\documentclass[aps,prl,twocolumn,showpacs,preprintnumbers,amsmath,amssymb,superscriptaddress]{revtex4-2}

\usepackage{graphicx}
\usepackage{dcolumn}
\usepackage{bm}
\usepackage{hyperref}
\hypersetup{
	colorlinks,
	citecolor=red,
	filecolor=black,
	linkcolor=blue,
	urlcolor=blue
}
\usepackage{lipsum}
\usepackage{natbib}

\usepackage{mathptmx}
\usepackage{dsfont}

\newcommand{\nc}{\newcommand}
\nc{\be}{\begin{equation}}
\nc{\ee}{\end{equation}}
\nc{\bal}{\begin{align}}
\nc{\eal}{\end{align}}
\nc{\bea}{\begin{eqnarray}}
\nc{\eea}{\end{eqnarray}}
\nc{\bean}{\begin{eqnarray*}}
\nc{\eean}{\end{eqnarray*}}
\nc{\mb}{\mbox}
\nc{\rnc}{\renewcommand}
\nc{\vk}{\mb{\bf k}}
\nc{\vp}{\mb{\bf p}}
\nc{\vn}{\mb{\bf n}}
\nc{\vq}{\mb{\bf q}}
\nc{\rr}{\mb{\bf r}}
\nc{\vz}{\hat {\mb{\bf z}}}
\nc{\vj}{\mb{\boldmath$j$}}
\nc{\vg}{\mb{\boldmath$g$}}
\nc{\x}{\mb{\boldmath$x$}}
\nc{\A}{\mb{\boldmath$A$}}
\nc{\va}{\mb{\boldmath$a$}}
\nc{\vs}{\mb{\boldmath$\sigma$}}
\nc{\vpi}{\mb{\boldmath$\pi$}}
\nc{\nab}{\nabla}
\nc{\X}{\sf x}
\nc{\kk}{{\bf k}}
\nc{\pp}{{\bf p}}
\nc{\qq}{{\bf q}}
\nc{\upspin}{{\uparrow}}
\nc{\dspin}{{\downarrow}}
\nc{\vecq}{{\bf q}}
\nc{\veck}{{\bf k}}
\nc{\vecp}{{\bf p}}
\nc{\vecl}{{\bf l}}
\nc{\vecr}{{\bf r}}
\nc{\vecx}{{\bf x}}
\nc{\vecR}{{\bf R}}
\nc{\vecG}{{\bf G}}
\nc{\vecA}{{\bf A}}
\nc{\vecpi}{{\bf \pi}}
\nc{\vecL}{{\bf L}}
\nc{\vecK}{{\bf K}}

\def\be{\begin{eqnarray}}
\def\ee{\end{eqnarray}}

\nc{\argg}{\text{Arg}}
\nc{\bd}{\textbf}
\nc{\bds}{\boldsymbol}
\nc{\ham}{\hat{\mathcal{H}}}
\nc{\im}{\text{Im}}
\nc{\la}{\langle}
\nc{\ra}{\rangle}
\nc{\re}{\text{Re}}
\nc{\rn}[1]{%
	\textup{\uppercase\expandafter{\romannumeral#1}}%
}
\nc{\sgn}{\text{Sgn}}
\nc{\tit}{\textit}
\nc{\tr}{\text{Tr}}

\nc{\les}{\leqslant}
\nc{\ges}{\geqslant}

\usepackage{stackengine}
\newcommand\xrowht[2][0]{\addstackgap[.5\dimexpr#2\relax]{\vphantom{#1}}}

\begin{document}

\title{Superfluid weight cross-over and critical temperature enhancement in singular flat bands}

\author{Guodong Jiang}
\affiliation{Department of Physics, University of Nevada, Reno, Reno NV 89502, USA}
\affiliation{Department of Applied Physics, Aalto University School of Science, FI-00076 Aalto, Finland}
\author{P{\"a}ivi T{\"o}rm{\"a}}
\affiliation{Department of Applied Physics, Aalto University School of Science, FI-00076 Aalto, Finland}
\author{Yafis Barlas}
\email{ybarlas@unr.edu}
\affiliation{Department of Physics, University of Nevada, Reno, Reno NV 89502, USA}




\begin{abstract}
Non-analytic Bloch eigenstates at isolated band degeneracy points exhibit singular behavior in the quantum metric. Here, a description of superfluid weight for zero-energy flat bands in proximity to other high-energy bands is presented, where they together form a singular band gap system. When the singular band gap closes, the geometric and conventional contributions to the superfluid weight as a function of the superconducting gap exhibit different cross-over behaviors. The scaling behavior of superfluid weight with the band gap is studied in detail, and the effect on the Berezinskii-Kosterlitz-Thouless (BKT) transition temperature is explored. It is discovered that tuning the singular band gap provides a unique mechanism for enhancing the supercurrent and critical temperature of two-dimensional (2D) superconductors.
\end{abstract}


\maketitle


{\em Introduction}.---Flat bands or regions in momentum space with a large density of states \cite{khodel1990superfluidity,kopnin2011high,kopnin2011surface,kopnin2013high,PhysRevLett.111.046604,peotta2015superfluidity}, like saddle point Van Hove singularities~\cite{vanhove1953,hirsch1986enhanced,volovik1994fermi,shaginyan2022} might be promising for the realization of high-$T_c$ superconductors. However, due to the vanishing band curvature in 2D perfectly flat bands, phase fluctuations can drive the critical temperature $T_c$ to near zero, even for strong pairing interactions \cite{emery1995}. One mechanism for this is the BKT transition, which occurs due to vortex proliferation at temperatures $T > T_c$, resulting in resistive vortex motion destroying superfluidity. The BKT transition temperature $T_c$ of 2D superconductors is determined from the Nelson-Kosterlitz criteria: $T_{c} =(\pi/8)D_s (T_c)$ where $D_s$ denotes the superfluid weight (superfluid stiffness) \cite{KTpaper1973,PhysRevLett.39.1201}, which defines the supercurrent $\bd{J}_s$ as a response to the vector potential $\bd{A}$ as $\bd{J}_s=-D_s\bd{A}$. Therefore, it is critical to identify band structure properties that can enhance the superfluid weight and establish conditions for achieving maximum values of $T_c$.

The superfluid weight in conventional BCS theory is inversely proportional to the effective mass, which vanishes for flat bands. This raises questions about the nature of superconductivity in moire superconductors hosting flat bands \cite{Cao2018,andrei2020graphene,balents2020sc,andrei2021marvels,kennes2021moire,torma2022sc}. In flat-band superconductors band geometric effects rescue the superconductivity, providing a finite superfluid weight \cite{peotta2015superfluidity,julku2016geometric,PhysRevB.94.245149,liang2017band,torma2018quantum,iskin2018exposing,PhysRevLett.123.237002,xie2020topology,PhysRevB.102.184504,PhysRevB.102.201112,PhysRevLett.127.170404,PhysRevLett.127.246403,torma2022superconductivity,PhysRevLett.123.237002,Tian2023,tanaka2024kinetic,PhysRevB.109.214518}. The superfluid weight of an isolated flat band is proportional to the integral of quantum metric, which is lower bounded by topological invariants. Therefore, band geometric properties determine the BKT transition temperature $T_c$ of flat-band superconductors~\cite{peotta2015superfluidity,julku2016geometric}. However, this analysis requires the flat band to be separated from other bands by a band gap $E_g$. In this Article, we study the effect of the band gap $E_g$ on the superfluid weight of flat bands in proximity to other bands. We  show its intriguing cross-over behavior as a function of the superconducting gap $\Delta$ when $E_g$ is varied in a class of flat bands called singular flat bands \cite{PhysRevB.78.125104,PhysRevB.99.045107,PhysRevLett.125.266403}. This behavior is due to the non-analyticity of Bloch functions in singular flat bands, which can be used to enhance $T_c$ of flat-band superconductors.

\begin{figure}[t]
\centering
\includegraphics[width=0.8\linewidth]{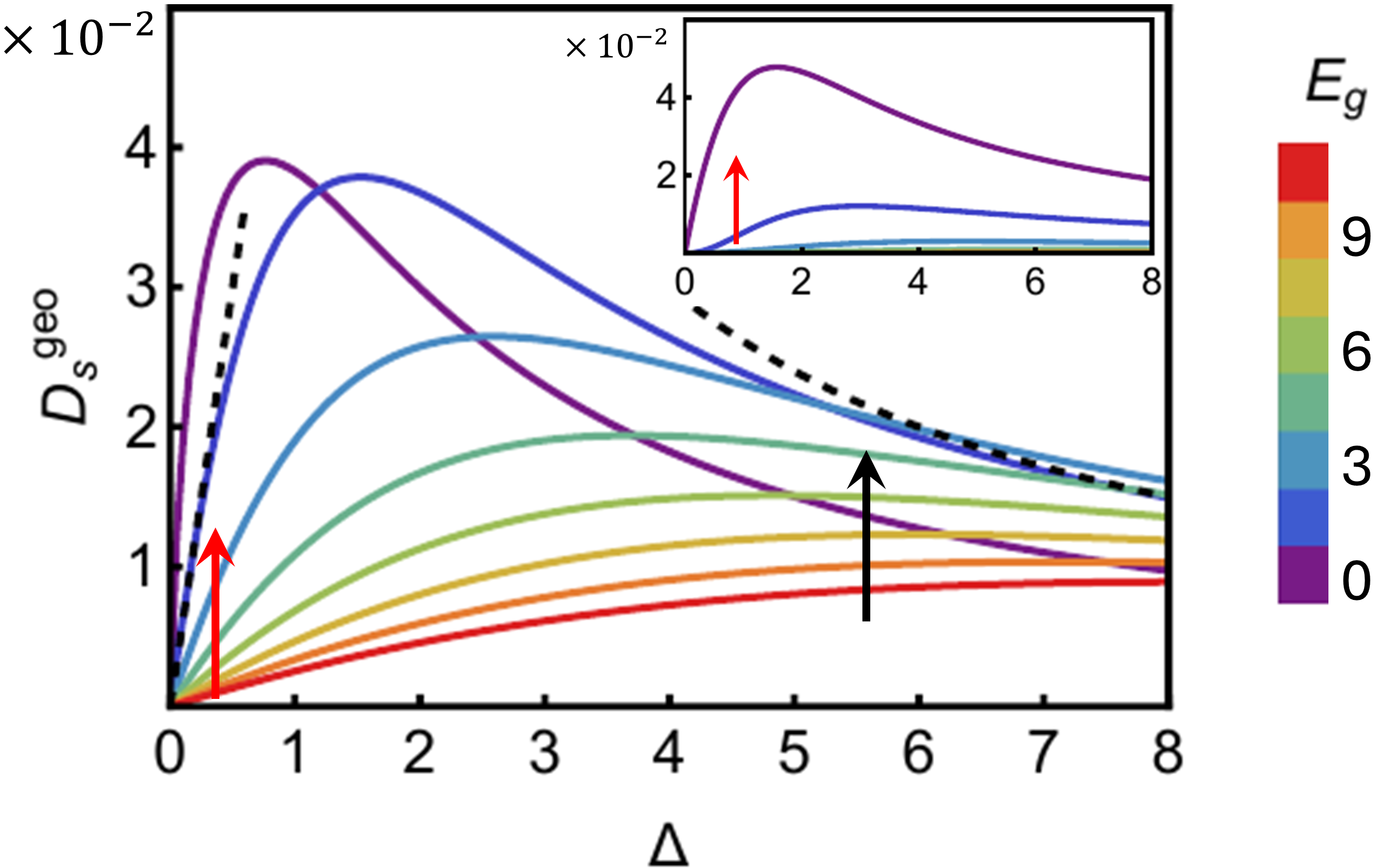}
\caption{Illustration of the cross-over of geometric superfluid weight $D_s^\text{geo}$ as a function of superconducting gap $\Delta$, in some fixed energy unit, as the band gap $E_g$ between a singular flat and dispersive band changes. In the weak interaction case (indicated by red arrow), as $E_g$ closes $D_s^\text{geo}$ cross-overs from $\propto\Delta$ to $\Delta\ln(W/2\Delta)$, with $W\approx 2$ the width of the dispersive band. The strong interaction case (black arrow) cross-overs from $\propto\Delta$ to $1/\Delta$. The two dashed curves represent the $\propto\Delta$ and $1/\Delta$ asymptotic behaviors for the blue curve $E_g=1.5$. Inset: the conventional superfluid weight $D_s^\text{conv}$ vs $\Delta$. In the weak interaction regime, it cross-overs from $\propto\Delta^2$ to $\Delta$. All the cross-overs above occur at the regime $E_g\sim\Delta$.}
\label{fig:cross-over}
\end{figure}

Singular flat bands exhibit essential (immovable) singularities in the Bloch eigenstates \cite{PhysRevB.99.045107} resulting from the destructive interference of Bloch phases due to band touching degeneracies \cite{PhysRevB.78.125104,PhysRevLett.125.266403}. Here, we introduce two classes of two- and three-band continuum Hamiltonians that host singular flat bands, separated from other high-energy bands of width $W_d$ by a tunable band gap $E_g$. Both classes of models have a momentum-space vortex with nonzero winding numbers at the gap closing point and are continuum model extensions of the Mielke~\cite{Mielke_1991} and Lieb~\cite{lieb1989two} lattices. For these lattice systems, it was found that $D_s$ scales as $ U\ln(\epsilon_c/U)$ due to the $k^{-2}$ divergence of the quantum metric near the band touching point \cite{julku2016geometric,iskin2019origin,wu2021superfluid,huhtinen2022revisiting}, where $U>0$ is the attractive interaction and $\epsilon_c$ is a constant. We show this logarithmic enhancement of $D_s$ is a general feature of singular flat bands. In the weak interaction limit, defined by $\Delta\ll W_d$, the geometric contribution to the superfluid weight, $D_s^\text{geo}$ exhibits a cross-over from $\Delta$ to $\Delta\ln(W/2\Delta)$ ($W$ is an energy scale of the order of $W_d$) as $E_g$ closes, with the cross-over at $E_g \sim \Delta$, as indicated by the red arrow in Fig~\ref{fig:cross-over}. In contrast, the conventional contribution $D_s^\text{conv}$ has a $\Delta^2$ to $\Delta$ cross-over, as shown in the inset of Fig \ref{fig:cross-over}. In the strong interaction limit $\Delta\gg W_d$, $D_s^\text{conv}$ is suppressed, while $D_s^\text{geo}$ has the cross-over from $\Delta$ to $1/\Delta$, indicated by the black arrow in Fig~\ref{fig:cross-over}. The different scaling limits of the weak and strong interactions result in a maximum superfluid weight, which can be used to enhance the BKT transition temperature.


{\em Chiral singular flat bands}.---We discuss two generic classes of two- and three-band continuum models hosting non-analytic Bloch eigenstates at isolated band degeneracy points. A minimal example of a singular gap system with a flat band in proximity to a dispersive band is the two-band continuum model for up-spins
\begin{align}
\label{eq:hchiral}
h(\bd{k})=& \zeta_J k^J[\cos(J\varphi_\bd{k})\hat{\sigma}_x + \sin(J\varphi_\bd{k})\hat{\sigma}_y]+m\hat{\sigma}_z \nonumber\\ 
&+(\zeta_J^2k^{2J}+m^2)^{1/2}  \hat{\sigma}_0,
\end{align}
where $\zeta_J$ is a constant with $J$ the integer chirality index, $m$ is the mass, and $k$, $\varphi_\bd{k}$ are the magnitude and azimuth angle of the 2D momentum. Pauli matrices $\hat{\sigma}_i$ act on the orbital degree of freedom of the model. We impose time-reversal symmetry and assume zero spin-orbit coupling, which dictates that $h(\bd{k})$ has a spin-down time-reversal partner. The spectrum of Eq.~\ref{eq:hchiral} includes a flat valence band $ \varepsilon_{v\bd{k}}=0$ and a dispersive conduction band $\varepsilon_{c\bd{k}}=2(\zeta_J^2k^{2J}+m^2)^{1/2}$. The bands acquire a winding number $\pm J$ upon opening the band gap $E_g=2m$ and are topological.

\begin{figure}[b!]
\centering
\includegraphics[width=0.48\textwidth]{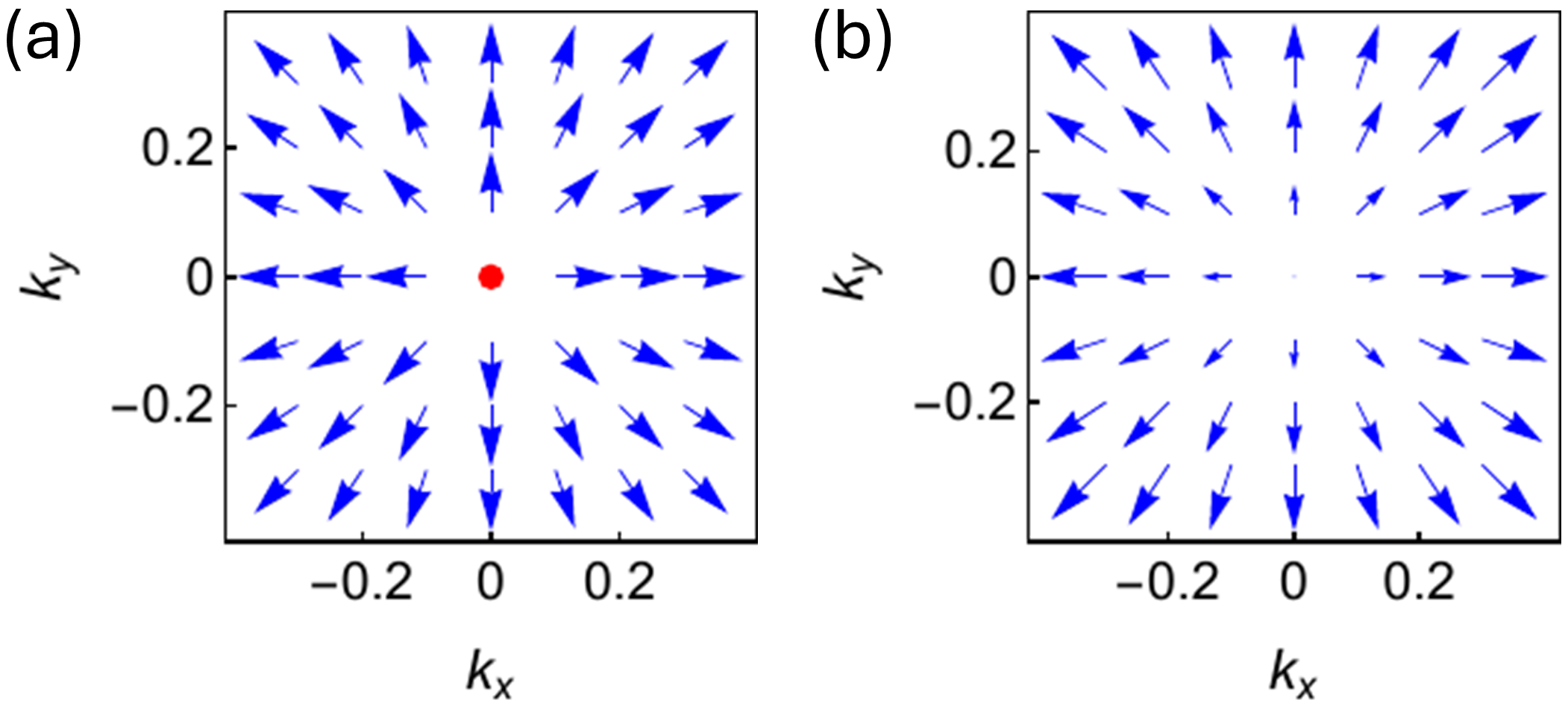}
\caption{Momentum-space vortex of the vector field $\hat{\bd {h}}(\kk)$ for model Eq. \ref{eq:hchiral} of $J=1$. (a) $m=0$, $\hat{\bd {h}}(\kk)$ is frustrated at the origin. (b) $m \neq 0$, $\hat{\bd {h}}(\kk)$ gets tilted out of the $k_x-k_y$-plane.}
\label{fig:bandtouch}
\end{figure}

Non-analyticity of Bloch eigenstates near the gap closing point can be visualized by plotting the unit vector field $\hat{{\bf h}}(\bd{k})={\bf h}(\bd{k})/|{\bf h}(\bd{k})|$, with $\bd{h}(\bd{k})=(h_x,h_y,h_z)$ the vector associated with matrices $\hat{\sigma}_x,\hat{\sigma}_y,\hat{\sigma}_z$ in Eq. \ref{eq:hchiral}. It exhibits a momentum-space vortex due to the winding number $J$, as indicated in Fig.~\ref{fig:bandtouch}. When $m=0$ (Fig.~\ref{fig:bandtouch} a), $\hat{{\bf h}}(\bd{k})$ is frustrated at the origin due to the non-analyticity of the Bloch states. However, when $m \neq 0$ (Fig.~\ref{fig:bandtouch} b), $\hat{{\bf h}}(\bd{k})$ cants out of the $\bd{k}$-plane at the origin, thus is well-defined. The non-analyticity is also reflected in the quantum metric of both bands \cite{hu2023effect}:
\begin{align}
\label{eq:twobandqm}
g_{\mu\nu}(\bd{k})=\frac{\zeta_J^2J^2k^{2J-2}}{4(\zeta_J^2k^{2J}+m^2)^2} \bigg[m^2\delta_{\mu\nu} +\frac{\zeta_J^2(k^2\delta_{\mu\nu}-k_\mu k_\nu)}{k^{2-2J}} \bigg].
\end{align}
At $m = 0$, the essential singularity at the origin leads to the singularity of quantum metric, $g_{\mu\nu} \propto k^{-2}$, whose divergence is independent of the chirality index $J$.

The two-band chiral model can be generalized to the three-band case, with Hamiltonian
\begin{align}
\label{eq:hchiralthree}
h(\bd{k})=C(m,k)\{\zeta_J k^J[\cos(J\varphi_\bd{k})\hat{\lambda}_x + \sin(J\varphi_\bd{k})\hat{\lambda}_y]+m\hat{\lambda}_z\},
\end{align}
where $\hat{\lambda}_i$'s form a three-dimensional representation of the SU(2) algebra. The additional factor $C(m,k)$ is to allow for generic dispersive bands while preserving the chirality~\footnote{Considering the fixed electron density, we engineer the band structure by $C(m,k)$ for the three-band model only. For the two-band model, when both bands are flattened, the chemical potential $\mu=0$ (aligned with the valence band) and electron density cannot be fixed simultaneously, making the analytical analysis more difficult. For the three-band model, there is no such limitation.}. We focus on two limiting cases: (i) $C(m,k)=1$, the model is finite-hopping and consists of a zero-energy flat band (0) and two dispersive bands ($\pm$), with $\varepsilon_{\pm,\bd{k}}= \pm (\zeta_J^2 k^{2J} + m^2)^{1/2}$; (ii) $C(m,k)=c(m)(\zeta_J^2k^{2J}+m^2)^{-1/2}$, the dispersive bands are flattened by long-range hoppings and the function $c(m)$ depicts how the band gap depends on the mass $m$.

For any choice of $C(m,k)$, model Eq. \ref{eq:hchiralthree} has both particle-hole and inversion symmetry, implying a zero inter-band (band-resolved) quantum metric between the valence and conduction band, $g^{+-}_{\mu\nu}(\bd{k})=0$, while the intra-band quantum metric $g^0_{\mu\nu},g^{\pm}_{\mu\nu}$ are all proportional to Eq.~\ref{eq:twobandqm} (see supplemental materials Sec. S3). The middle flat band is topologically trivial, in contrast to the flat band of the two-band model. In case (i) when $C(m,k)=1$, the model of $J=1$ corresponds to the continuum limit of Lieb lattice, with a topological mass gap $m$ from a loop-ordered state \cite{PhysRevB.85.155106}. It is distinct from the bipartite Lieb lattice model discussed in Ref.~\cite{julku2016geometric} where a band gap occurs due to unequal hoppings between the nearest neighbors. The relation of the two- and three-band models above to the singular flat bands previously studied in finite-hopping lattice models \cite{PhysRevB.78.125104,PhysRevB.99.045107,rhim2021singular,PhysRevLett.125.266403} is provided in the supplemental materials Sec. S1.

{\em Superfluid Weight}.---To analyze the energy cost of superconducting phase fluctuations in singular flat bands, we calculate the superfluid weight for onsite attractive interactions. For the models we study in this Article, the orbitals are spatially separated, therefore onsite attractive interactions lead to intra-orbital interactions of the form $\hat{V}=-\sum_\alpha U_\alpha \hat{n}_{i\alpha\uparrow}\hat{n}_{i\alpha\downarrow}$, where $i$ labels the unit cell, $\alpha$ labels the orbital, and $\hat{n}_{i\alpha\sigma}=c_{i\alpha\sigma}^\dagger c_{i\alpha\sigma}$ is the density operator for spin $\sigma$. Then mean-field decoupling results in an intra-orbital order parameter: $\Delta_\alpha=-U_\alpha\la c_{i\alpha\downarrow}c_{i\alpha\uparrow}\ra$. With the uniform pairing condition $\Delta_{\alpha}\equiv\Delta $ for all orbitals $\alpha$ in the singular flat bands, the total superfluid weight $D_{s} = D_{s}^\text{conv} + D_s^\text{geo}$ for the two-band chiral model can be expressed as
\begin{align}
\label{eq:totalsw}
D_s= \sum_{\kk} \frac{\Delta^2}{E_{c,\kk}^3} (\partial_x \varepsilon_{c,\kk} )^2 + 4 \Delta \sum_{\kk} \bigg( 1 - \frac{\Delta}{E_{c,\kk}} \bigg) g_{xx}(\kk),
\end{align}
where $E_{c,\kk} =(\varepsilon_{c,\kk}^2 + \Delta^2)^{1/2}$ is the quasiparticle energy, and the sum of $\bd{k}$ is limited by the bandwidth $W_d$ (for derivations see supplementary materials Sec. S3). We have used the relation $D_{s,xx} = D_{s,yy}$ and $D_{s,xy}=0$ to define $ D_{s} = D_{s,xx}$. Similar expressions of superfluid weight for the three-band chiral model with particle-hole and inversion symmetry can also be derived (see supplemental materials Sec. S3.1 for details). For some cases, we nominally adjusted the interactions $U_{\alpha}$ to ensure the uniform pairing condition.

\begin{figure}[t!]
\centering
\includegraphics[width=0.49\textwidth]{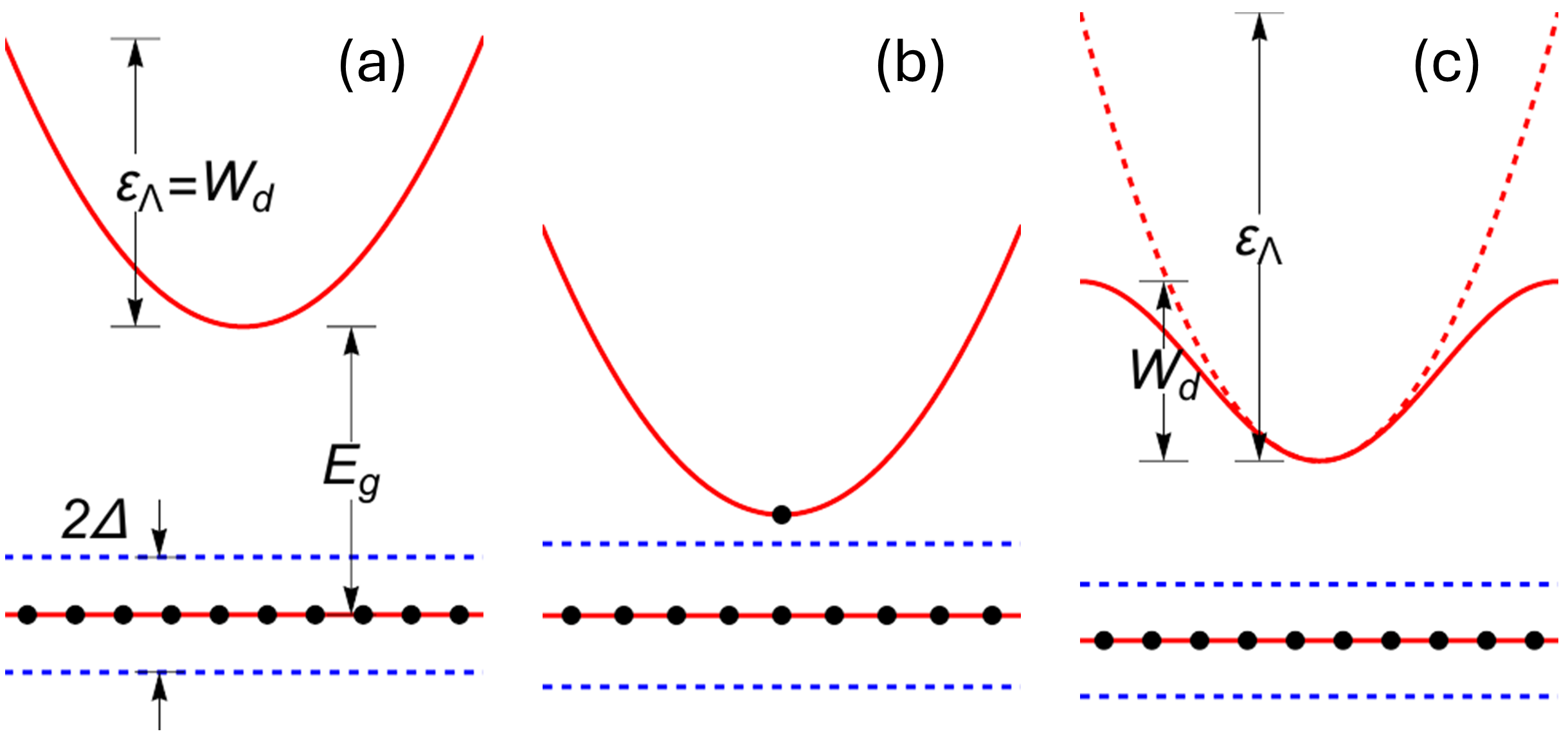}
\caption{Schematic illustration of the gap closing between a singular flat and dispersive band. The three energy scales are the band gap $E_g$, the superconducting gap $\Delta$, and the cutoff energy $\varepsilon_\Lambda$ (defined by a cutoff momentum $k_\Lambda$) or the dispersive bandwidth $W_d$. The electron density is fixed as the flat band at half-filling. Blue dashed lines indicate the interaction window $2\Delta$ around the Fermi level. (a)(b), for continuum models, $\varepsilon_\Lambda=W_d$. (a) At $E_g\gg\Delta$, $\mu=0$ lies in the flat band. (b) As $E_g$ approaches $\Delta$, the dispersive band gets electron-doped, and both $\mu,\Delta$ renormalize, but the renormalization of $\mu$ is neglected. (c), for lattice models, the flat band may span the entire Brillouin zone, with $\varepsilon_\Lambda$ interpreted as the intercept of dispersion interpolation at the zone edge, and $\varepsilon_\Lambda\neq W_d$.}
\label{fig:gap}
\end{figure}

{\em Weak interaction limit}.---We first study the weak interaction limit ($\Delta \ll W_d$) of superfluidity for a system with flat bands at half-filling ($\mu \approx 0$) in proximity to other dispersive bands, as illustrated in Fig. \ref{fig:gap}. The system is characterized by three energy scales: the band gap $E_g$, the superconducting gap $\Delta$, and the cutoff energy $\varepsilon_\Lambda$ or width $W_d$ of the dispersive band. The effect of other energy scales, e.g., the width $W_f$ of the flat band with weak dispersion, is assumed to be subleading and neglected. The superconducting gap is characterized by a single parameter $\Delta$, consistent with the uniform pairing condition \cite{peotta2015superfluidity}, i.e., all orbitals have the same pairing potential. We envision that the band gap $E_g$ can be tuned at a fixed electron density for the superconducting state. When $E_g\gg\Delta$ (Fig. \ref{fig:gap}(a)), the dispersive band has zero electron density. As $E_g$ approaches $\Delta$ (Fig. \ref{fig:gap}(b)) from above (below), the dispersive band becomes electron- (hole-) doped, shifting the chemical potential downwards (upward), $\delta\mu<0$ ($>0$). In the weak interaction regime $\Delta\ll \varepsilon_\Lambda$ or $W_d$, the change $\delta\mu$ is neglected as a first approximation. Since the density of states changes, the order parameter of each orbital $\Delta_\alpha$ ($\alpha$ is the orbital index) renormalizes with $E_g$. However, for chiral models, a moderate change of $E_g$ still corresponds to an approximate uniform pairing, allowing us to take the average of $\Delta_\alpha$ as $\Delta$ (see supplemental materials Sections S5, S6, S7 for details).

We propose a general expression for the zero-temperature superfluid weight of such systems when the singular flat band is at half-filling. It can be argued from scaling analysis and the non-analyticity of Bloch functions; therefore, it is model-independent. It contains two pieces: the conventional contribution from the dispersion of dispersive bands \cite{schrieffer1999theory,tinkham2004introduction}, $D_s^\text{conv}$, and the geometric contribution from the intra- and inter-band effects of all bands \cite{peotta2015superfluidity,liang2017band}, $D_s^\text{geo}$. The lattice geometric effect \cite{huhtinen2022revisiting,chan2022pairing,tam2024geometry,PhysRevB.109.214518} is neglected for our continuum models. Both terms can be written in a universal form
\begin{align}
\label{eq:generic}
D_{s}^{\alpha}=\Delta \big[ \mathcal{F}^\alpha_1(\lambda,\kappa) \chi(\lambda,\kappa) +\mathcal{F}_2^\alpha(\lambda,\kappa) \big],
\end{align}
where $\mathcal{F}_1^\alpha,\mathcal{F}_2^\alpha$ are functions of dimensionless parameters $\lambda \equiv E_g/W$ and $\kappa \equiv W/\Delta$, with $W$ a phenomenological bandwidth of the dispersive band. Label $\alpha=``\text{conv}"$ or $``\text{geo}"$ denotes the two contributions, and $\chi(\lambda,\kappa)$ is a cross-over function, defined as
\begin{align}\label{eq:cross-over}
\chi(\lambda,\kappa)\equiv \ln \bigg(\frac{1+\sqrt{1+\kappa^2}}{1+\sqrt{1+\lambda^2\kappa^2}}\bigg).
\end{align}
$\lambda$ controls the band geometry of the gap system, whereas the quantity $\lambda\kappa=E_g/\Delta$ measures the magnitude of $E_g$ vs $\Delta$, determining the cross-over.

We now explain the cross-over from Eq.~\ref{eq:generic} above for the weak interaction regime $\Delta\ll W$ ($\kappa\gg1$). When the band gap is large, $E_g \gg \Delta$ ($\lambda\kappa\gg1$, Fig. \ref{fig:gap}(a)), $\kappa$ cancels in the logarithmic leading term of the cross-over function so $\chi(\lambda,\kappa)\approx-\ln\lambda+...$, leading to the expansion of Eq. \ref{eq:generic} in powers of $1/\kappa$:
\begin{align}
\label{eq:power}
D_s^\alpha=\Delta\bigg[f_1^\alpha(\lambda)+f_2^\alpha(\lambda)\frac{1}{\kappa}+...\bigg],
\end{align}
where $f_1^\alpha,f_2^\alpha$ are the two leading coefficients. It shows that $D_s^\text{geo}$ and $D_s^\text{conv}$ scale as power laws of $\Delta$ when the flat band is isolated. Typically $f_1^\text{geo}\neq 0$ while $f_1^\text{conv}=0,f_2^\text{conv}\neq0$ for perfectly flat bands, implying the behavior $D_s^\text{geo}\propto \Delta$ and $D_s^\text{conv}\propto\Delta^2$, respectively. Here $D_s^\text{conv}$ is of higher order because the pairing at the edge of the dispersive bands is extremely weak.

As the band gap closes such that $E_g\sim\Delta$ ($\lambda\kappa\sim 1$, Fig. \ref{fig:gap}(b)), $\kappa$ in the logarithmic leading term of $\chi(\lambda,\kappa)$ survives. In particular, when the gap completely closes ($\lambda=0$), $\chi(0,\kappa) \approx \ln (\kappa/2)$ leading to
\begin{align}
\label{eq:log}
D_s^\alpha\approx\Delta\bigg[\mathcal{F}_1^\alpha(0,\kappa)\ln\frac{\kappa}{2}+\mathcal{F}_2^\alpha(0,\kappa)\bigg].
\end{align}
This gives a $D_s^\alpha\propto\Delta\ln(W/2\Delta)$ behavior if $\mathcal{F}_1^\alpha(0,\kappa) \neq0 $, leading to a power law to logarithm cross-over. However, if $\mathcal{F}_1^\alpha(0,\kappa)=0$, then Eq. \ref{eq:log} corresponds to another power law $D_s^\alpha\propto\Delta\mathcal{F}_2^\alpha(0,\kappa)$, and $\lambda\kappa\sim1$ is the cross-over regime between two power laws. We find $\mathcal{F}_1^\text{conv}(0,\kappa)=0$, leading to a $\Delta^2$ to $\Delta$ cross-over for $D_s^\text{conv}$; whereas $\mathcal{F}_1^\text{geo}(0,\kappa)\neq0$ due to the $k^{-2}$ divergence of quantum metric, leading to the distinctive $\Delta$ to $\Delta\ln(W/2\Delta)$ cross-over for $D_s^\text{geo}$ in singular flat bands.

\begin{table}[b!]
\centering
\caption{\label{tab:functiontab} Functions $\mathcal{F}_i^\alpha$ and the leading coefficients $f_i^\alpha$ of the two-band chiral model Eq.\ref{eq:hchiral}.}
\begin{tabular}{c|c}
\hline
\xrowht{12pt}
$\mathcal{F}_1^\text{geo}$&$\frac{J}{8\pi}(2-\lambda^2\kappa^2)$ \\
\hline
\xrowht{12pt}
$\mathcal{F}_2^\text{geo}$& $\frac{J}{8\pi}(1-\lambda^2-\lambda^2\kappa^2\ln\lambda-\sqrt{1+\lambda^2\kappa^2}+\lambda^2\sqrt{1+\kappa^2})$\\
\hline
\xrowht{12pt}
$f_i^\text{geo}$& $f_1^\text{geo}=\frac{J}{8\pi}(1-\lambda^2-2\ln\lambda)$\\
\hline
\xrowht{12pt}
$\mathcal{F}_1^\text{conv}$& $\frac{J}{4\pi}\lambda^2\kappa^2$\\
\hline
\xrowht{12pt}
$\mathcal{F}_2^\text{conv}$& $\frac{J}{4\pi}\big[\lambda^2\kappa^2\ln\lambda+(1+\lambda^2\kappa^2)(\frac{1}{\sqrt{1+\lambda^2\kappa^2}}-\frac{1}{\sqrt{1+\kappa^2}})\big]$\\
\hline
\xrowht{12pt}
$f_i^\text{conv}$& $f_1^\text{conv}=0$,\,\,\,$f_2^\text{conv}=\frac{J}{4\pi}(\frac{\lambda^2}{3}-1+\frac{2}{3\lambda})$\\
\hline
\end{tabular}
\end{table}

Next, we address the physics of the two-band chiral model. Due to the combination of particle-hole and inversion symmetry, at half-filling, the superfluid weight of the three-band model in the dispersive case (i) exhibits the same functional form as the two-band model (see supplemental materials Sec. S3.1 for details); therefore, the discussion below applies to it as well. Functions $\mathcal{F}^\alpha_i$, $f^\alpha_i$ of the two-band model are listed in Table~\ref{tab:functiontab}. Here, the phenomenological bandwidth is defined as $W\equiv(\varepsilon_\Lambda^2+E_g^2)^{1/2}$, with the cutoff energy $\varepsilon_\Lambda=2\zeta_Jk_\Lambda^J$ and band gap $E_g=2m$. The chirality index $J$ is a multiplying factor, indicating that the cross-over is universal for singular band-gap systems of different chiralities. The cross-over can also be seen from the divergence of the leading coefficients $f_1^\text{geo}$ and $f_2^\text{conv}$ in Table \ref{tab:functiontab} as $\lambda\rightarrow0$---if they did not diverge, then no cross-over would occur. Coefficients $f_1^\text{geo}$ and $f_2^\text{conv}$ diverge as $\ln\lambda$ and $1/\lambda$, respectively, pointing to a nontrivial character of the geometric superfluid weight, which will be discussed shortly.

Although we focused on the half-filling case, the cross-over behavior persists at other fillings of flat bands as long as $\nu$ is not too close to 0 or 1. Away from half-filling, one must account for a nonzero $\mu$, which introduces another energy scale. Since the cross-over occurs whenever the edge of the dispersive bands approaches the interaction window of $2\Delta$ around the chemical potential, it defines $|E_g-\mu| \sim \Delta$  as the cross-over regime. However, flat bands' electron or hole density must dominate for this. If $\nu$ is very close to 0 or 1, the chemical potential may lie in the dispersive bands, in which case the analysis above does not apply.

\begin{figure}[b!]
\centering
\includegraphics[width=0.48\textwidth]{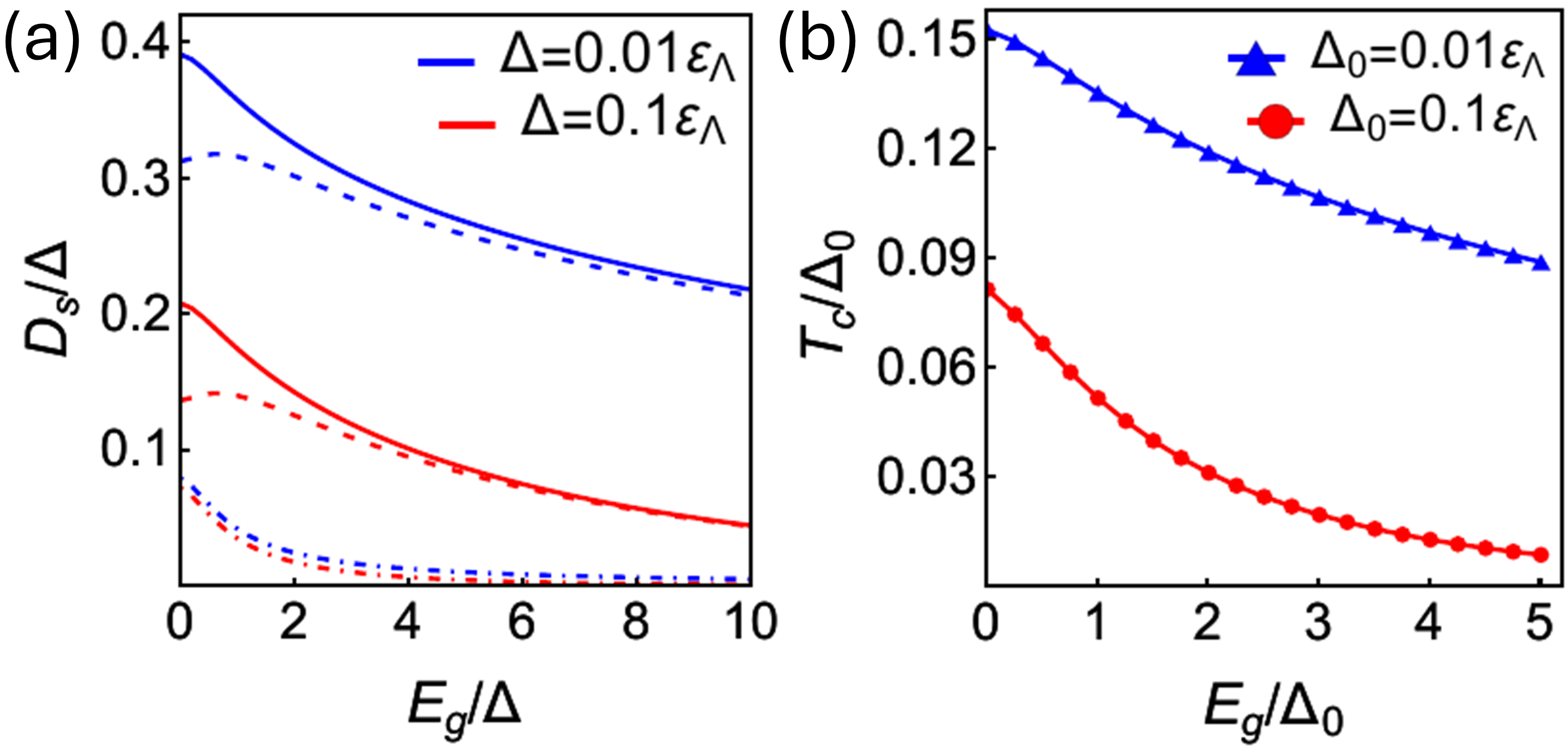}
\caption{(a) $D_s/\Delta$ vs $E_g/\Delta$ scaling plot and (b) BKT $T_c$ for the two-band model Eq. \ref{eq:hchiral}, with $J=1$, $E_g=2m$ and $\varepsilon_\Lambda=2\zeta_Jk_\Lambda^J$. (a) shows the total ($D_s$, solid), geometric ($D_s^\text{geo}$, dashed) and conventional ($D_s^\text{conv}$, dot-dashed) contributions to the superfluid weight. In (b), $U_{A,B}$ are fixed at $0.99\varepsilon_\Lambda$ (blue) and $8.51\varepsilon_\Lambda$ (red), giving the uniform pairing $\Delta_0=0.01\varepsilon_\Lambda$ (blue) and $0.1\varepsilon_\Lambda$ (red) at $E_g=T=0$, respectively.}
\label{fig:2band}
\end{figure}

{\em Scaling behavior of superfluid weight vs band gap}.---The dimensionless function $D_s/\Delta$ can be used to quantify the phase fluctuations; therefore, it is helpful to study the scaling behavior of $D_s/\Delta$ vs $E_g/\Delta$, where the latter controls the cross-over. In Fig. \ref{fig:2band}(a), we plot it at two fixed $\Delta$ values for the two-band model for $J=1$. Using Eq. \ref{eq:generic}, $D_s/\Delta$ can be expressed in terms of independent variables $\lambda\kappa$ and $\kappa$, where we find the change $\delta D_s^\alpha/\Delta$ has single scaling dependence on $\lambda\kappa$ and is insensitive to $\kappa$ value, i.e., all curves of the same type are parallel in Fig. \ref{fig:2band}(a). This is a general property of the weak interaction limit and is independent of the flat-band filling. The two contributions $D_s^\text{geo}$ and $D_s^\text{conv}$ also show distinct scaling behaviors: $D_s^\text{geo}$ exhibits a weak logarithmic enhancement in the regime $E_g\gg \Delta$, but saturates around $E_g \sim \Delta$ due to the short-range nature of the cross-over (see supplemental materials Sec. S4). In contrast, $D_s^\text{conv}$ only begins to grow after the dispersive bands become electron- or hole-doped at $E_g \sim \Delta$.

{\em BKT enhancement}.---When the band gap closes, the ratio $D_s/\Delta >(J/4 \pi) \ln (\varepsilon_\Lambda/2\Delta)$ diverges weakly as $\Delta\rightarrow0$, indicating that for extremely weak interactions $T_c$ can be driven close to the mean-field temperature $T_{MF}$ by the singular gap closing. Calculating $D_s$ at finite temperatures and applying the Nelson-Kosterlitz criteria, we plot the BKT $T_c$ as a function of $E_g$ for the two-band model for $J=1$ at two weak couplings in Fig. \ref{fig:2band}(b). Different from the $D_s/\Delta$ vs $E_g/\Delta$ scaling plot, here we include the slight renormalization of $\Delta$ due to the change of density of states by the incoming dispersive band. We choose the superconducting gap at $T=E_g=0$, $\Delta_0$ as the energy unit since it approximately scales with the interaction $U$. This choice is more unbiased than the fixed unit $\varepsilon_\Lambda$ because then $T_c/\Delta_0$ contains the information of how efficiently the superconducting gap is converted to $T_c$. The $T_c$ enhancement for the large $\Delta_0$, large $E_g$ case (red curve) is larger than the other (blue curve) because its flat band has much weaker quantum geometry at $E_g/\Delta_0=5$. We also note that in the second phase of the gap closing process ($E_g/\Delta_0<1$), the enhancement comes from $D_s^\text{conv}$ of the dispersive band; the geometric effect, although more singular, is canceled by the inter-band effect \cite{julku2016geometric,PhysRevB.109.214518}.

{\em Strong interaction limit}.---We now discuss the strong interaction limit ($\Delta\gg W_d$) of the gap closing effect on the superfluid weight by analyzing case (ii) of the three-band model. We compare gap closing of a singular and non-singular flat band within the same model at $J=1$, by choosing $c(m)=m$ for the former and $c(m)=c_0$ for the latter, where $E_g=c_0$ then is independent of $m$. For convenience, we also fix $m$ to be $m_0=0.42\varepsilon_\Lambda$ for the non-singular model. At this value, the three orbitals have equal weight in all three bands, so the uniform pairing condition is satisfied. A non-singular band gap closing has the general property that the quantum metric $g_{\mu\nu}$ may vary with $\bd{k}$ and the band gap $E_g$, but never diverges when $E_g$ completely closes. It implies that the function $g_{\mu\nu}(\bd{k},E_g)$ is bounded, which justifies the choice of $c(m)=c_0$ for the non-singular case. For the flattened model, the conventional superfluid weight vanishes, therefore we only discuss the geometric superfluid weight below.

\begin{figure}[t!]
\includegraphics[width=0.48\textwidth]{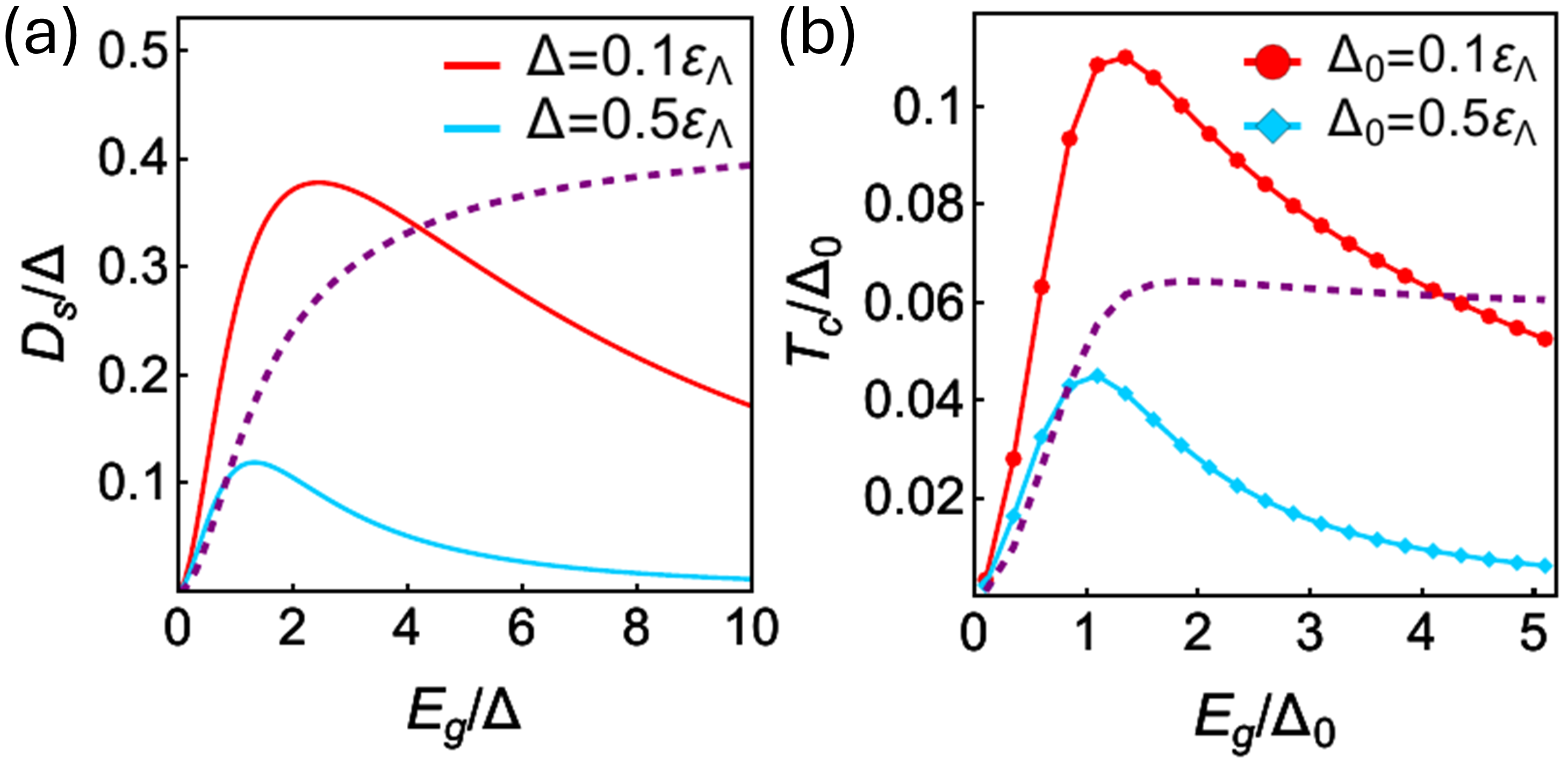}
\caption{Contrasting the singular (solid) and non-singular (dashed) gap closing, using the flattened model Eq. \ref{eq:hchiralthree} with $J=1$, $E_g=c(m)$ and $\varepsilon_\Lambda=\zeta_Jk_\Lambda^J$. Singular case: $c(m)=m$. Non-singular case: $c(m)=c_0$, then fix $m=0.42\varepsilon_\Lambda$. (a) $D_s/\Delta$ vs $E_g/\Delta$ scaling plot. (b) BKT $T_c$ plot, with $U_{A,B,C}$ fixed at $2.51\varepsilon_\Lambda$ (red) and $12.56\varepsilon_\Lambda$ (light blue), so the uniform pairing at $E_g=T=0$ is $\Delta_0=0.1\varepsilon_\Lambda$ and $0.5\varepsilon_\Lambda$, respectively. The non-singular curves in (a)(b) are independent of $\Delta$ or $\Delta_0$ value and are uniformly shown in purple.}
\label{fig:3bandf}
\end{figure}

Since the width of the dispersive bands is zero, the nature of cross-over for $D_s^\text{geo}$ is also changed, with a cross-over function $\chi_f(\lambda\kappa)=1-(1+\lambda^2\kappa^2)^{-1/2}$ instead of Eq. \ref{eq:cross-over} ($\chi_f$ can be derived from Eq. \ref{eq:totalsw} by taking the strong interaction limit). $\chi_f$ depends on the single parameter $\lambda\kappa=E_g/\Delta$, so $D_s^\text{geo}\propto \Delta\chi_f$ exhibits a $\propto\Delta$ to $1/\Delta$ cross-over in the regime $E_g\sim\Delta$. This is expected for the strong interaction case when the interaction scale includes all the bands and the superfluid weight vanishes as $1/\Delta$. What distinguishes the singular gap closing is the $D_s/\Delta$ vs $E_g/\Delta$ scaling plot, as shown by the solid curves in Fig. \ref{fig:3bandf}(a), where a maximum is achieved at $E_g/\Delta\sim 1$ and then decreases to zero at $E_g=0$, in contrast to a monotonic decrease for the non-singular gap closing (dashed curve). In the strong interaction case, $\Delta$ is enhanced by the gap closing more strongly than in the weak interaction case, leading to a competition between $\Delta$ and $D_s$ as $E_g$ changes, which determines $E_g\sim\Delta$ as the optimal condition for $T_c$ of 2D superconductors. In Fig. \ref{fig:3bandf}(b), we plot the BKT transition temperature $T_c$ for both the singular and non-singular cases. The singular case has a sharp peak at $E_g\sim\Delta$, since in the regime $E_g>\Delta$, both $\Delta$ and $D_s$ increase as $E_g$ decreases. On the contrary, although $\Delta$ is enhanced, $T_c$ of the non-singular case is almost unchanged at $E_g>\Delta$ due to the decrease of $D_s/\Delta$. As $E_g/\Delta$ gets smaller than 1, $T_c$ of both cases decreases rapidly due to the inter-band cancellation effect for $D_s$ \cite{julku2016geometric,PhysRevB.109.214518}. This cancellation ($D_s\rightarrow 0$, as $E_g\rightarrow0$) behavior is due to $E_{c,\bd{k}}\rightarrow\Delta$ in the second term of Eq. \ref{eq:totalsw} when the dispersive bandwidth becomes smaller than $\Delta$ over the entire Brillouin zone.

{\em Results for large chirality $J$}.---As $J$ increases, the band geometry gets stronger; the dispersive bands exhibit a $k^J$-dispersion when the gap closes \footnote{These $J$-chiral models resemble the effective model of $J$-layer rhombohedral stacked multi-layer with only nearest-neighbor inter-layer hoppings. In general, remote hopping effects lead to the trigonal warping of bands and instability for large values of $J$, resulting in smaller values of $J$-chiral models at neighboring points in the BZ.}. As a result, the $D_s/\Delta$ axis of the $D_s/\Delta$ vs $E_g/\Delta$ scaling plots for all models (Fig. \ref{fig:2band}(a) and \ref{fig:3bandf}(a)) will scale with $J$, like the two-band case shown in Table \ref{tab:functiontab}, leading to the possibility of $D_s/\Delta\gg1$ for large $J$. Then, the phase fluctuation is weak ($T_c/T_{MF}\approx 1$), and $T_c$ is limited by the superconducting gap $\Delta$ rather than $D_s$. In other words, too strong band geometry and dispersion weaken the role of $D_s$ enhancement through singular band gap closing. For completeness, we present the results for the $J=6$ case in the supplementary materials Sec. S8. For large values of $J$ with strong interactions, non-singular gap closing has the same sharp $T_c$ enhancement peak as the singular gap closing since both models experience the same enhancement of $\Delta$ due to the change in the density of states (see Fig. S8(b) in the supplementary materials, cf. Fig. \ref{fig:3bandf}(b)).

{\em Discussion}.---The mechanism to enhance the critical temperature in quasi-flat band superconductors is not unique. Recent studies have found that the conventional superfluid weight in quasi-flat bands can be enhanced by interband pair-exchange interactions~\cite{paramasivam2024high}. However, nontrivial band geometry is a fundamental property of flat bands that cannot originate from spatially decoupled atomic orbitals. We showed that the interplay between the band gap $E_g$ and superconducting gap $\Delta$ determines the cross-over behavior of the superfluid weight in singular flat bands, which is purely geometric in origin. Similar behavior is expected for systems with concentrated quantum metric that scales as $g_{\mu\nu}(\bd{k}) \propto k^{-2}$~\cite{herzog2024topological}. In finite-hopping tight-binding models, singular flat bands originating from quantum interference of Bloch phases \cite{PhysRevB.78.125104,PhysRevLett.125.266403} are also characterized by non-analytic Bloch eigenstates~\cite{PhysRevB.99.045107,rhim2021singular}. In this case, for the two-band model, the flat band acquires a dispersion upon adding a mass $m$, with the flat bandwidth $W_f$ proportional to $m$, whereas it remains dispersionless in the three-band model. For these systems with $W_f\sim m$, the geometric superfluid weight will be approximately $\Delta\ln(W_d/\text{max}\{W_f,\Delta\})$ during the gap closing. Our analysis also reveals that the supercurrent and BKT transition temperature is enhanced and optimized by gap closing in a singular flat band system, paving the road for searching for high-$T_c$ flat-band superconductors.



{\em Acknowledgement.}---The authors would like to thank Andrew J. Millis and Grigory E. Volovik for helpful discussions. This work was supported by the award number DE-SC0022178 from the Department of Energy (Y.B.), the UNR/VPRI Startup funds PG19012 (Y.B \& G.J.) and the Research Council of Finland (former Academy of Finland) under project number 339313 and by Jane and Aatos Erkko Foundation, Keele Foundation, and Magnus Ehrnrooth Foundation as part of the SuperC collaboration (P.T. \& G.J.).






%

\pagebreak
\onecolumngrid

\setcounter{figure}{0}
\setcounter{equation}{0}
\renewcommand\thefigure{S\arabic{figure}}
\renewcommand\theequation{S\arabic{equation}}

\section*{Supplementary Information for ``Superfluid weight cross-over and critical temperature enhancement in singular flat bands"}
\section{S1: Flat band touching in short-range hopping models}
In general, due to the locality of the projection operator, a topological perfectly flat band is not allowed in finite-range models ~\cite{Chen_2014}. The Bloch functions of a group of isolated degenerate flat bands in short-range hopping models are smooth analytic nonvanishing functions of Bloch phases defined globally on the Brillouin zone (BZ) torus. Therefore, they are topologically trivial, i.e., they must have zero total Chern number~\cite{Chen_2014}. The Berry curvature and quantum metric are smooth analytic functions in this case. The above arguments only apply to finite range hopping models where the Hamiltonian can be expressed as a finite sum of Bloch phases. This is, of course, not true for flat bands originating from continuum models, such as Landau levels and twisted 2D crystals or infinite range hopping models. More generally, the vector bundle associated with a gapped spectrum of filled bands with compact localized states is polynomial and analytic, hence topologically trivial~\cite{PhysRevB.95.115309}. Compact localized states (CLS) (not to be confused with maximally localized Wannier functions) have finite amplitude only in a bounded region in real space. However, nonanalytic Bloch wavefunctions do show up in gapless flat bands from short-range hopping Hamiltonian and are topologically nontrivial \cite{PhysRevB.95.115309,PhysRevB.92.205307}. 

Gapless flat band spectrum, which consists of a flat band touching a dispersive band, can be designed from line graphs and split graphs of bipartite lattices~\cite{PhysRevB.78.125104,PhysRevLett.125.266403}. In lattice models, this situation occurs from destructive interference of Bloch phases, where the band touching degeneracy originates from real space topology~\cite{PhysRevB.78.125104,PhysRevB.99.045107}. While all CLS live within the flat band, they are incomplete; the flat band also accommodates linearly independent real-space extended eigenstates \cite{PhysRevB.78.125104}. This type of band touching is classified as singular, as it is associated with immovable singularities of the Bloch function \cite{PhysRevB.99.045107,rhim2021singular}. In the case of finite-range hopping models, the flat band acquires a dispersion when the degeneracy is lifted. Flat band lattice models can also exhibit non-singular band degeneracies, which are also associated with smooth analytic nonvanishing functions globally defined over the BZ torus.

Finally, we would like to remark on the two-band model Eq. 1 in the main text (MT). In Eq. 1, the $\hat{\sigma}_0$ term coefficient cannot be expressed as a polynomial of $k_x,k_y$. Therefore, it introduces infinite-range hoppings, which makes the valence band perfectly flat. However, when the chirality index $J$ is even, the $m=0$ case of Eq. 1 can be realized in finite-range models, as the coefficient of $\hat{\sigma}_0$ becomes $\zeta_Jk^J$ which is a polynomial of $k_x,k_y$ ($J=2$ corresponds to Kagome and Mielke lattice singular flat bands~\cite{PhysRevB.78.125104,iskin2019origin,PhysRevB.99.045107,PhysRevLett.125.266403}). Similarly, one can check that for the three-band model Eq. 3 the unflattened case (i) can always be realized in finite-range models, whereas the flattened case (ii) cannot be realized with a local Hamiltonian.

\section{S2: Superfluid Weight Calculation for Singular Two-Band Models}
\label{app:dscalculation}
We use the following expression for the superfluid weight under the uniform pairing condition \cite{PhysRevB.109.214518},
\begin{align}\label{eq:dst}
D_{s,\mu\nu}(T)=&\sum_{l,\bd{k}} \frac{\Delta^2}{E_{l\bd{k}}^3}\bigg(\tanh\frac{\beta E_{l\bd{k}}}{2}-\frac{\beta E_{l\bd{k}}}{2}\text{sech}^2\frac{\beta E_{l\bd{k}}}{2}\bigg)\partial_\mu \xi_{l\bd{k}}\partial_\nu\xi_{l\bd{k}} \nonumber\\
&+\sum_{l,\bd{k}} \tanh\frac{\beta E_{l\bd{k}}}{2}\bigg\{\frac{4\Delta^2}{E_{l\bd{k}}}g^l_{\mu\nu}(\bd{k})+\sum_{l'\neq l}8\Delta^2\bigg[\frac{p^{(+)}_{ll'}(\bd{k})}{E_{l\bd{k}}+E_{l'\bd{k}}}+\frac{p^{(-)}_{ll'}(\bd{k})}{E_{l\bd{k}}-E_{l'\bd{k}}}\bigg]g^{ll'}_{\mu\nu}(\bd{k})\bigg\},
\end{align}
the $T=0$ limit of which is
\begin{align}\label{eq:ds0}
D_{s,\mu\nu}(T=0)=\sum_{l,\bd{k}}\bigg[\frac{\Delta^2}{E_{l\bd{k}}^3}\partial_\mu \xi_{l\bd{k}}\partial_\nu\xi_{l\bd{k}}+\frac{4\Delta^2}{E_{l\bd{k}}}g^l_{\mu\nu}(\bd{k})\bigg]+\sum_{l'>l,\bd{k}}\frac{16\Delta^2p^{(+)}_{ll'}(\bd{k})}{E_{l\bd{k}}+E_{l'\bd{k}}}g^{ll'}_{\mu\nu}(\bd{k}).
\end{align}
This is consistent with the general expression in Ref. \cite{peotta2015superfluidity,liang2017band}. Here $l,l'$ are band indices, $\Delta$ is the uniform pairing order parameter either at finite temperature $T=1/\beta$ or $T=0$, and $E_{l\bd{k}}=\sqrt{\xi_{l\bd{k}}^2+\Delta^2}$ is the quasiparticle energy, with $\xi_{l\bd{k}}=\varepsilon_{l\bd{k}}-\mu$. The first and second line of Eq. \ref{eq:dst} correspond to the conventional and geometric contributions to superfluid weight, respectively.
\begin{align}
\label{eq:pfactor}
p^{(\pm)}_{ll'}(\bd{k})=\frac{1}{2}\bigg(1\pm\frac{\xi_{l\bd{k}}\xi_{l'\bd{k}}+\Delta^2}{E_{l\bd{k}}E_{l'\bd{k}}}\bigg),\,\,\,l\neq l',
\end{align}
are the inter-band coherence factors and
\begin{align}\label{eq:glldef}
&g^l_{\mu\nu}(\bd{k})=\re\la \partial_{\mu} u_{l\bd{k}} |(1- |u_{l\bd{k}} \ra \la u_{l\bd{k}} |)|\partial_{\nu} u_{l\bd{k}} \ra, \nonumber\\
&g^{ll'}_{\mu\nu}(\bd{k})=-\re\la\partial_\mu u_{l\bd{k}}|u_{l'\bd{k}}\ra\la u_{l'\bd{k}}|\partial_\nu u_{l\bd{k}}\ra,\,\,\,l\neq l'.
\end{align}
are the intra- and inter-band (band-resolved) quantum metric tensor.

In the following, we present the $T=0$ superfluid weight calculation for singular flat bands in proximity to dispersive bands. $\mu =0$ is fixed, so the flat bands are approximately half-filling. For continuum models, azimuthal symmetry implies $D_{s,xx} = D_{s,yy} = D_s$ and $D_{s,xy}=0$, so only the $D_{s,xx}$ component is calculated.
\subsection{S2.1: Two-band Chiral Model}
\label{app:twobandchiral}
For the two-band model Eq. 1 in MT, since the valence band remains flat when the band gap is open, the conventional contribution of superfluid weight only comes from the conduction band. At $T=0$, it can be expressed as
\be 
\label{eq:twobandconv}
D_{s}^\text{conv} = \sum_\bd{k} \frac{\Delta^2 }{E_{c\kk}^3} (\partial_x \varepsilon_{c\kk})^2,
\ee
with $ \varepsilon_{c\kk} = 2 \sqrt{\zeta_J^2 k^{2J} + m^2}$ the conduction band energy. Evaluating Eq. \eqref{eq:twobandconv} with a momentum cutoff $ k_\Lambda $ gives 
\begin{align}
D_s^\text{conv}=\frac{J \Delta}{4 \pi} \bigg[\lambda^2 \kappa^2 \bigg(\ln \frac{1+ a_{\kappa}}{1+a_{\lambda\kappa}}+\ln \lambda \bigg)+a_{\lambda \kappa} - \frac{a_{\lambda \kappa}}{a_{\kappa}}\bigg], 
\end{align}
where $a_{x} \equiv \sqrt{1 + x^2}$ and we naturally introduced $\lambda=m/\sqrt{\zeta_J^2 k_\Lambda^{2J} + m^2}$ and $\kappa=2 \sqrt{\zeta_J^2 k_\Lambda^{2J} + m^2}/\Delta$.

For the geometric superfluid weight, one can show for any two-band model with $\varepsilon_{v\kk} =\mu= 0$, there is $D_s^{\text{geo},vc} = -2D_s^{\text{geo},c}$, so the total geometric superfluid weight can be expressed as $D^\text{geo}_s =  D_s^{\text{geo},v} - D_{s}^{\text{geo},c}$, giving
\be 
\label{eq:twobandgeo}
D_{s}^{\text{geo}} = 4 \Delta \sum_{\kk}  \bigg( 1 - \frac{\Delta}{E_{c\kk}} \bigg) g_{xx} (\kk).
\ee
Eq.~\eqref{eq:twobandgeo} can be evaluated using the quantum metric expression Eq. 2 in MT. We find the valence and conduction band geometric superfluid weights are
\begin{align}
D^{\text{geo},v}_s = \frac{J \Delta}{8 \pi}( 1- \lambda^2 - 2 \ln \lambda),
\end{align}
\begin{align}
D_s^{\text{geo},c} = \frac{J \Delta}{8 \pi} \bigg[( \lambda^2 \kappa^2 -2) \bigg( \ln  \frac{1+ a_{\lambda \kappa}}{1+a_{\kappa}} +\ln \lambda \bigg)+ a_{\lambda \kappa} - \lambda^2 a_{\kappa} \bigg].
\end{align}
Notice the $-2\ln\lambda $ term in the two equations above cancel each other, leading to a logarithmic term $\lambda^2\kappa^2\ln\lambda$, which is regular as $\lambda\rightarrow0$.

\subsection{S2.2: Dirac-point Touching}
\label{app:dirac}
The Dirac Hamiltonian
\begin{align}\label{eq:dirac}
h(\bd{k})=v(k_x\hat{\sigma}_x+k_y\hat{\sigma}_y),
\end{align}
with $v$, the velocity has a quantum metric for both bands
\begin{align}\label{eq:diracqm}
g_{\mu\nu}(\bd{k})=\frac{k^2\delta_{\mu\nu}-k_\mu k_\nu}{4k^4}.
\end{align}
This model has no band gap or flat band, but we now show how the physics is related.

Aligning the chemical potential with the Dirac point, $\mu=0$, the problem has two energy scales $vk_\Lambda$ and $\Delta$, with $k_\Lambda$ the cutoff momentum, from which only one dimensionless parameter $\kappa=vk_\Lambda/\Delta$ can be defined. The zero-temperature superfluid weight can be found using Eq. \eqref{eq:ds0}, giving the conventional and geometric superfluid weight
\begin{align}\label{eq:normaldiracconv}
D_s^{\text{conv}}=\frac{\Delta}{\pi}\bigg(1-\frac{1}{a_\kappa}\bigg)=2D_s^\text{geo},
\end{align}
which are proportional to $\Delta$ in the weak coupling limit.

To exhibit a $\Delta\ln(a/\Delta)$ behavior, we assume model \eqref{eq:dirac} lies in the linear regime of some fictitious parameter $\lambda$. To incorporate with Eq. 4 in MT, we allow the group velocity of the two bands to be different by adding a $\sigma_0$ term, giving the modified Hamiltonian
\begin{align}\label{eq:modifieddirac}
h(\bd{k})=v(k_x\hat{\sigma}_x+k_y\hat{\sigma}_y)+v'k\hat{\sigma}_0.
\end{align}
The two bands have group velocity $v_c=v'+v$ and $v_v=v'-v$, respectively, but the quantum metric is still Eq. \eqref{eq:diracqm}. Now the model has three energy scales $v_vk_\Lambda$, $v_ck_\Lambda$ and $\Delta$, from which we define $\lambda=-v_v/v_c$ and $\kappa=v_ck_\Lambda/\Delta$. We assume $v'\geqslant 0$, so $|\lambda|\leqslant1$. At the critical value $\lambda=0$, the lower band becomes flat.

Eq. \eqref{eq:ds0} now gives the total geometric superfluid weight
\begin{align}\label{eq:diracds}
D_s^{\text{geo}}=\frac{\Delta}{4\pi}\frac{1+\lambda}{1-\lambda}\ln\frac{1+a_\kappa}{1+a_{\lambda\kappa}},
\end{align}
which contains the cross-over function $\chi(\lambda,\kappa)$, while the conventional superfluid weight can be shown not containing $\chi(\lambda,\kappa)$. Functions of this model are listed in Table \ref{tab:diractab}.

\begin{table}[ht]
\caption{\label{tab:diractab} Functions of the Dirac model \eqref{eq:modifieddirac}}
\begin{tabular}{c|c}
\hline
\xrowht{12pt}
$\mathcal{F}_1^\text{geo}$& $\frac{1}{4\pi}\frac{1+\lambda}{1-\lambda}$\\
\hline
\xrowht{12pt}
$\mathcal{F}_2^\text{geo}$& 0\\
\hline
\xrowht{12pt}
$f_1^\text{geo}$& $-\frac{1}{4\pi}\frac{1+\lambda}{1-\lambda}\ln|\lambda|$\\
\hline
\xrowht{12pt}
$\mathcal{F}_1^\text{conv}$& 0\\
\hline
\xrowht{12pt}
$\mathcal{F}_2^\text{conv}$& $\frac{1}{2\pi}\big(2-\frac{1}{\sqrt{1+\kappa^2}}-\frac{1}{\sqrt{1+\lambda^2\kappa^2}}\big)$\\
\hline
\xrowht{12pt}
$f_1^\text{conv}$& $\frac{1}{\pi}$\\
\hline
\end{tabular}
\end{table}

In the table both $f_1^\text{geo}$ and $f_1^\text{conv}\neq0$, indicating that for $\lambda\neq0$, in the regime $\lambda\kappa\gg 1$ both $D_s^\text{geo}$ and $D_s^\text{conv}\propto\Delta$. As the valence band becomes flat ($\lambda\rightarrow 0$), $D_s^\text{conv}$ remains linear while $D_s^\text{geo}$ has a $\Delta\sim \Delta\ln(v_ck_\Lambda/2\Delta)$ cross-over. Analyzing the slope function $f_1^\text{geo}$, at $\lambda=1$ it is finite, corresponding to the original Dirac model \eqref{eq:dirac}; $\lambda=0$ is the critical point when the valence band becomes flat, so it diverges; $\lambda=-1$ is a ``non-geometric limit" when the two bands become degenerate so it vanishes.
\section{S3: Superfluid Weight Calculation for Singular Three-Band Models}
\label{app:threeband}
\subsection{S3.1: Three-band Chiral model, case (i)}
\label{app:threebandchiral}
We choose the representation
\begin{align}
\hat{\lambda}_x =\begin{pmatrix}
0 & 1 & 0\\
1 & 0 & 0\\
0 & 0 & 0
\end{pmatrix},\,\,\, 
\hat{\lambda}_y =\begin{pmatrix}
0 & 0 & 0\\
0 & 0 & 1\\
0 & 1 & 0
\end{pmatrix},\,\,\, 
\hat{\lambda}_z =\begin{pmatrix}
0 & 0 & -i\\
0 & 0 & 0\\
i & 0 & 0
\end{pmatrix}.
\end{align}
for Eq. 3 in MT. A realization of this Hamiltonian for $J=1$ in Lieb lattice is shown in Fig. \ref{fig:lieblattice}(a). The imaginary hoppings $\pm im/4$ between $A$ and $C$ atoms open a band gap $m$ between the flat and valence/conduction band. $\hat{\lambda}_z$ term makes the model non-bipartite and Lieb theorem no longer applicable \cite{lieb1989two}, but one can still assume a superconducting exact ground state for small $m$.

Using Fourier transform $c_{\bd{R}\alpha}=\frac{1}{\sqrt{N}}\sum_\bd{k}e^{i\bd{k}\cdot(\bd{R}+\bd{r}_\alpha)}c_{\bd{k}\alpha}$, with $\bd{R}=i\hat{\bd{x}}+j\hat{\bd{y}}$ the lattice vectors, $\bd{r}_A=(\frac{1}{2},0)$, $\bd{r}_B=(0,0)$, $\bd{r}_C=(0,\frac{1}{2})$ the atom positions within a unit cell, the model in Fig. \ref{fig:lieblattice}(a) gives the Bloch Hamiltonian
\begin{align}\label{eq:hklieb}
h_{lattice}^{J=1}(\bd{k})=\begin{pmatrix}
   0 & 2t\cos\frac{k_x}{2} & -im\sin\frac{k_x}{2}\sin \frac{k_y}{2}\\
   2t\cos\frac{k_x}{2} & 0 & 2t\cos\frac{k_y}{2}\\
   im\sin\frac{k_x}{2}\sin \frac{k_y}{2} & 2t\cos\frac{k_y}{2} & 0
\end{pmatrix}.
\end{align}
Setting $\zeta_1=-t$ and expanding it around the gap closing point $\bd{k}_M=(\pi,\pi)$, one obtains Eq. 3 of MT for $J=1$.

One can show that PHS combined with inversion symmetry leads to an important property of the three-band chiral Hamiltonian Eq. 3, that the inter-band quantum metric between the valence and conduction bands vanishes, $g^{+-}_{\mu\nu}(\bd{k})=0$ (for definition see Eq. \eqref{eq:glldef}).

Consider a general Hamiltonian $h(\bd{k})$ with simultaneous PHS and inversion symmetry. The PHS is given by $U h^*(\bd{k}) U^{\dagger} = -h(-\bd{k})$ while inversion is given by $Vh(\bd{k})V^\dagger=h(-\bd{k})$, with $U,V$ both unitary matrices. Let $\pm$ denote the conduction and valence band related by this PHS. Then as long as $UV^* U^*=V$ holds, one can make a gauge choice for states $u_\pm(\bd{k})$ such that
\begin{align}\label{eq:foureq}
&|u_\mp(-\bd{k})\ra=U|u_\pm(\bd{k})^*\ra, \nonumber\\
&|u_\pm(-\bd{k})\ra=V|u_\pm(\bd{k})\ra,
\end{align}
where $|u^*\ra$ denotes the column vector with each component of $|u\ra$ taking complex conjugate. Properties $U$ and $V$ in this equation imply that the connection between valence and conduction band, $\mathcal{A}_\mu(\bd{\bd{k}})\equiv\la u_+(\bd{k})|\partial_\mu u_-(\bd{k})\ra$ is simultaneously an odd and even function of $\bd{k}$, i.e. by PHS
\begin{align}
\mathcal{A}_\mu(-\bd{k})=&\la u_+(-\bd{k})|\partial_\mu u_-(-\bd{k})\ra \nonumber\\
=&\la U u_-(\bd{k})^*|\partial_\mu [U u_+(\bd{k})^*]\ra \nonumber\\
=&\la u_-(\bd{k})^*|\partial_\mu u_+(\bd{k})^*\ra \nonumber\\
=&\la \partial_\mu u_+(\bd{k})|u_-(\bd{k})\ra=-\mathcal{A}_\mu(\bd{k}),
\end{align}
where to get the third line, we used the unitarity and $\bd{k}$-independence of $U$ matrix; similarly, by inversion symmetry
\begin{align}
\mathcal{A}_\mu(-\bd{k})=&\la u_+(-\bd{k})|\partial_\mu u_-(-\bd{k})\ra \nonumber\\
=&\la Vu_+(\bd{k})|\partial_\mu [Vu_-(\bd{k})]\ra \nonumber\\
=&\la u_+(\bd{k})|\partial_\mu u_-(\bd{k})\ra=\mathcal{A}_\mu(\bd{k}).
\end{align}
$\mathcal{A}_\mu(\bd{\bd{k}})$ being both odd and even leads to $\mathcal{A}_\mu(\bd{\bd{k}})=0$. This result can be further shown to be gauge-invariant due to the orthogonality between states $u_+(\bd{k})$ and $u_-(\bd{k})$, therefore $g^{+-}_{\mu\nu}(\bd{k})=0$.

Inspecting Hamiltonian Eq. 3 in MT, we find it satisfies PHS $U=\text{diag}(1,1,1)$ and inversion $V=\text{diag}(-1,1,-1)$ for odd $J$, while $U=\text{diag}(1,-1,1)$ and $V=\text{diag}(1,1,1)$ for even $J$. Therefore $g^{+-}_{\mu\nu}(\bd{k})=0$ is justified.

These symmetries also lead to: (1) Berry curvature of the valence and conduction band are exactly opposite, while the flat band has no Berry curvature. (2) quantum metric of the three bands $g^0_{\mu \nu},g^{\pm}_{\mu \nu}$ are related to the inter-band (band-resolved) quantum metric in a way $g^{0}_{\mu \nu} (\bd{k})  = 2 g^{\pm}_{\mu \nu} (\bd{k})= - 2 g^{0 \pm}_{\mu \nu}(\bd{k})$. The analytical calculation can show $g^0_{\mu \nu}$ is precisely four times the quantum metric of the two-band model, Eq. 2 in MT.

Using Eq. \eqref{eq:ds0}, we find the total geometric superfluid weight is
\begin{align}
\label{eq:dsgeo3band}
D_{s}^{\text{geo}} = 4 \Delta \sum_{\kk}  \bigg( 1 - \frac{\Delta}{E_{+,\kk}} \bigg) g^0_{xx}(\kk),
\end{align}
where $E_{+,\kk}=\sqrt{\varepsilon_{+,\bd{k}}^2+\Delta^2}$, with $\varepsilon_{+,\bd{k}}=\sqrt{\zeta_J^2 k^{2J} + m^2}$. Compared with Eq. \eqref{eq:twobandgeo}, the conventional and geometric superfluid weights are similar to the two-band chiral model. The slight difference is here we define $E_g=m$, $W=\sqrt{\zeta_J^2k_\Lambda^{2J}+m^2}$ and $\lambda=E_g/W$, $\kappa=W/\Delta$. Then, one can check the functions $\mathcal{F}_i^\alpha,f_i^\alpha$ are also similar to those in Table 1 of MT, except for some numerical factors.
\subsection{S3.2: Three-band Chiral model, case (ii)}
For case (ii) of model Eq. 3, the geometry is the same as case (i). Eq. \eqref{eq:dsgeo3band} is still valid, but with $\varepsilon_{+,\bd{k}}=c(m)$ since the valence and conduction bands are flattened. Performing the integrals, one obtains
\begin{align}
D_s^\text{geo}=\frac{J\Delta}{2\pi}\bigg(1-\frac{1}{\sqrt{1+t^2}}\bigg)(-2\ln \lambda_m+1-\lambda_m^2),
\end{align}
where $\lambda_m=m/\sqrt{\zeta_J^2k_\Lambda^{2J}+m^2}$ and $t=c(m)/\Delta$.
\subsection{S3.3: Lieb Lattice Model without Inversion Center}
\label{app:lieb}
\begin{figure}
\includegraphics[width=0.35\textwidth]{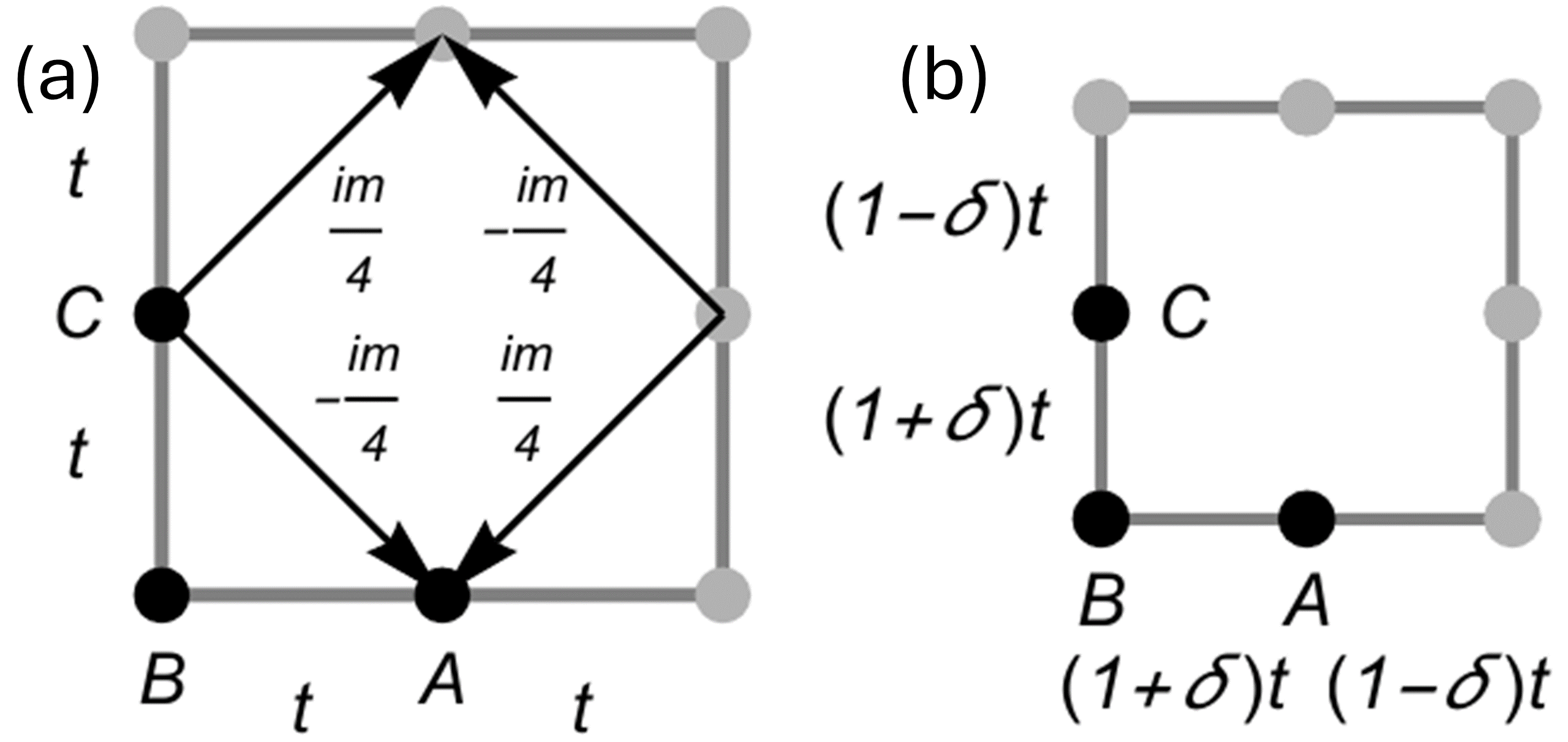}
\caption{Contrasting two types of Lieb lattice model with a band gap. Black dots denote the three atoms $A,B,C$ grouped as a unit cell. (a) A non-bipartite lattice with a gap $m$ opened by imaginary hoppings between atom $A$ and $C$. The arrows indicate hopping integrals from $C$ to $A$, while the reverse directions are their complex conjugate. (b) A bipartite lattice from Ref. \cite{julku2016geometric}, with a gap $2\sqrt{2}t\delta$ opened by unequal hoppings $(1\pm\delta)t$ between $A,B$ or $B,C$. (a) has inversion symmetry, but (b) does not.}
\label{fig:lieblattice}
\end{figure}

We also present the calculation for the Lieb lattice model in Ref. \cite{julku2016geometric} (Fig. \ref{fig:lieblattice}(b)). The Hamiltonian is
\begin{align}\label{eq:lieb}
h(\bd{k})=2t\begin{pmatrix}
0&a_\bd{k}^*&0\\
a_\bd{k}&0&b_\bd{k}\\
0&b_\bd{k}^*&0
\end{pmatrix}
\end{align}
where $a_\bd{k}=\cos(k_x/2)+i\delta\sin(k_x/2)$ and $b_\bd{k}=\cos(k_y/2)+i\delta\sin(k_y/2)$. Unlike the chiral model, it can be written as a linear combination of four SU(3) algebra generators.

The two dispersive bands of this model have energy $\varepsilon_{\pm,\bd{k}}=\pm 2t\sqrt{|a_\bd{k}|^2+|b_\bd{k}|^2}$. A gap $2\sqrt{2}t\delta$ to the flat band is opened at $\bd{k}_M=(\pi,\pi)$ when $\delta>0$. The continuum limit near $\bd{k}_M$ is
\begin{align}
\varepsilon_{\pm,\bd{k}}\approx\pm t\sqrt{8\delta^2+ (1-\delta^2) k^2},
\end{align}
where $\bd{k}$ is the momentum shift from $\bd{k}_M$ and the lattice spacing is $a=1$. The conduction and valance bands are related by PHS, $Uh(\bd{k})^*U^\dagger=-h(-\bd{k})$ with $U=\text{diag}(1,-1,1)$, but there is no inversion symmetry, giving a nonzero inter-band quantum metric $g^{+-}_{\mu\nu}(\bd{k})$. We find near $\bd{k}_M$
\begin{align}\label{eq:qmliebpm}
g^{+-}_{\mu\nu}(\bd{k})\approx -\frac{\delta^2(2\delta_{\mu\nu}-1)}{4(4\delta^2+\frac{1-\delta^2}{2}k^2)^2},
\end{align}
\begin{align}\label{eq:qmlieb0pm}
g^{0\pm}_{\mu\nu}(\bd{k})\approx-\frac{1}{8(4\delta^2+\frac{1-\delta^2}{2}k^2)^2}
\begin{pmatrix}
4\delta^2+(1-\delta^2)k^2-(1-\delta^4)k_x^2&4\delta^2-(1-\delta^2)^2k_xk_y\\
4\delta^2-(1-\delta^2)^2k_xk_y&4\delta^2+(1-\delta^2)k^2-(1-\delta^4)k_y^2
\end{pmatrix},
\end{align}
In the limit $\delta\rightarrow0$ there is relation $g^{+-}_{\mu\nu}(\bd{k})\propto \delta^2(\bd{k})$ and $g^{0\pm}_{\mu\nu}(\bd{k})\propto(k^2\delta_{\mu\nu}-k_\mu k_\nu)/k^4$. The continuum limit of the superfluid weight expression is complicated but follows the general expression Eq. 4 in MT.

For this model define $E_g=2\sqrt{2}t\delta$, $\varepsilon_\Lambda=t\sqrt{(1-\delta^2)k_\Lambda^2}$, $W=\sqrt{\varepsilon_\Lambda^2+E_g^2}$, then as usual $\lambda=E_g/W$, $\kappa=W/\Delta$ and keep in mind that $\delta$ is dimensionless. It turns out to be convenient to keep $\delta$ in the functions, and $\mathcal{F}_i^\alpha(\lambda,\kappa,\delta)$ and $f_i^\alpha(\lambda,\kappa,\delta)$ are listed in Table \ref{tab:liebtab}. We find the functions for the conventional superfluid weight are similar to those of the two-band chiral model (Table 1 of MT) if set $J=1/(1-\delta^2)$, and the cross-over behaviors are similar too.

\begin{table}[h!]
\caption{\label{tab:liebtab} Functions of the Lieb lattice model}
\begin{tabular}{c|c}
\hline
\xrowht{12pt}
$\mathcal{F}_1^\text{geo}$& $\frac{1}{2\pi}\big(2-\frac{1+\delta^2}{1-\delta^2}\lambda^2\kappa^2\big)$\\
\hline
\xrowht{30pt}
$\mathcal{F}_2^\text{geo}$& \begin{tabular}{c}
$\frac{1}{2\pi}\big[-\frac{1+\delta^2}{1-\delta^2}\lambda^2\kappa^2\ln\lambda+\frac{\lambda^2\kappa^2}{1-\delta^2}\big(\delta^2\frac{\sqrt{1+\kappa^2}-1}{\kappa^2}$\\
$-\delta^2\frac{\sqrt{1+\lambda^2\kappa^2}-1}{\lambda^2\kappa^2}+\frac{1}{\sqrt{1+\kappa^2}}-\frac{1}{\sqrt{1+\lambda^2\kappa^2}}\big)\big]$
\end{tabular}\\
\hline
\xrowht{12pt}
$f_i^\text{geo}$& $-\frac{1}{\pi}\ln\lambda$ \\
\hline
\xrowht{12pt}
$\mathcal{F}_1^\text{conv}$& $\frac{1}{4\pi}\frac{\lambda^2\kappa^2}{1-\delta^2}$\\
\hline
\xrowht{12pt}
$\mathcal{F}_2^\text{conv}$& $\frac{1}{4\pi}\frac{1}{1-\delta^2}\big[\lambda^2\kappa^2\ln|\lambda|+(1+\lambda^2\kappa^2)\big(\frac{1}{\sqrt{1+\lambda^2\kappa^2}}-\frac{1}{\sqrt{1+\kappa^2}}\big)\big]$ \\
\hline
\xrowht{12pt}
$f_i^\text{conv}$& $f_1^\text{conv}=0$,\,\,\,$f_2^\text{conv}=\frac{1}{4\pi}\frac{1}{1-\delta^2}(\frac{\lambda^2}{3}-1+\frac{2}{3|\lambda|})$\\
\hline
\end{tabular}
\end{table}
\section{S4: Short-range Nature of the $\Delta\sim\Delta\ln(W/2\Delta)$ cross-over}
\label{app:shortrange}
The $\Delta\sim\Delta\ln(W/2\Delta)$ cross-over for the geometric superfluid weight has a general short-range property, as explained below.

\begin{figure}[t!]
\includegraphics[width=0.48\textwidth]{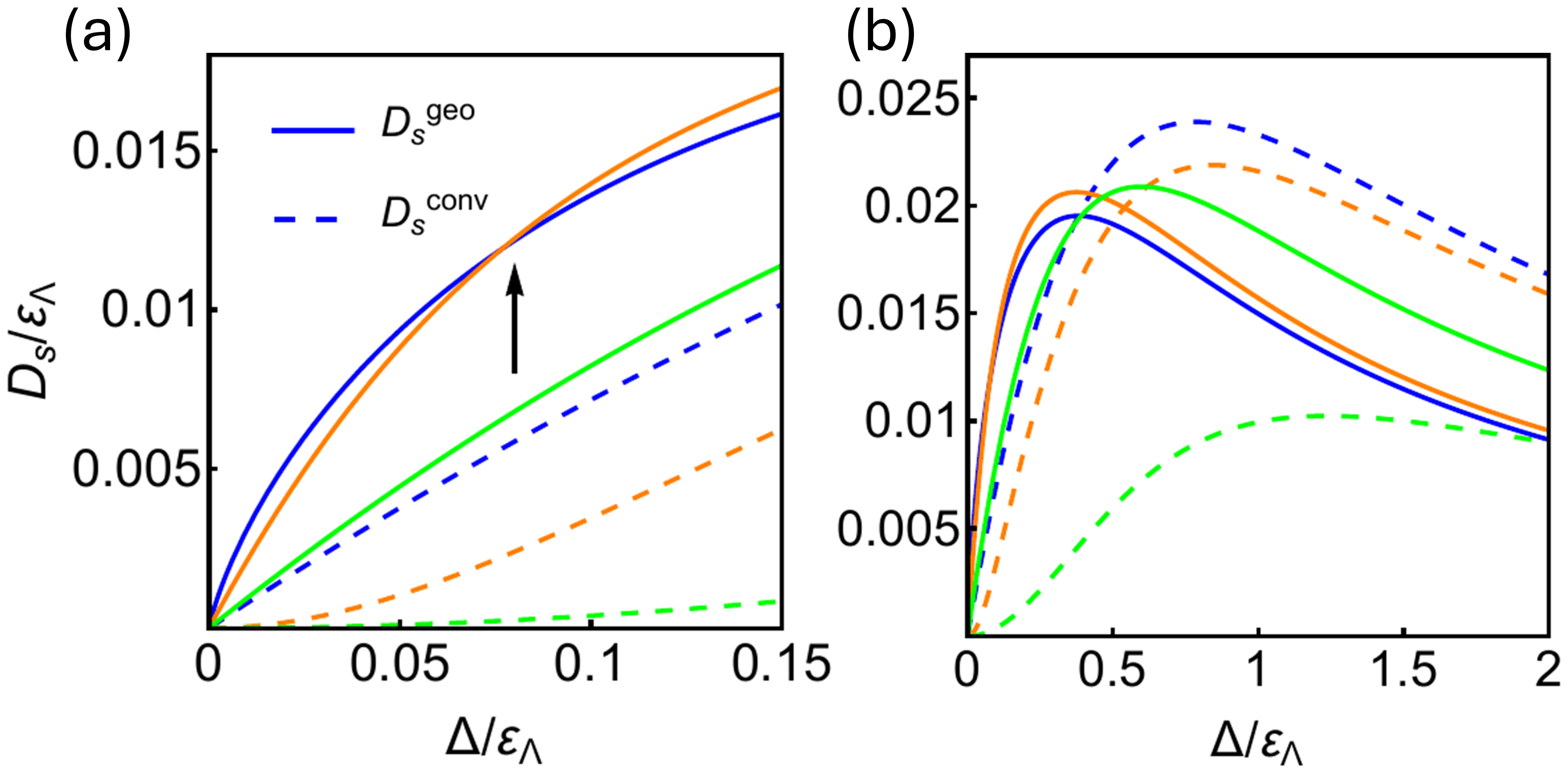}
\caption{$D_s$ vs $\Delta$ plot for the two-band model (using Eq. 4 and Table 1 of MT), for $m=0$ (blue), 0.1 (orange), 0.5 (green). The parameters are $\zeta_J=k_\Lambda=J=1$ so $\varepsilon_\Lambda=2\zeta_Jk_\Lambda^J=2$. (a) A smaller scale $0\leqslant\Delta\leqslant 0.15\varepsilon_\Lambda$. As the gap closes, the geometric superfluid weight has a $\Delta\sim\Delta\ln(a/\Delta)$ cross-over, while the conventional superfluid weight has a $\Delta^2\sim\Delta$ cross-over. The arrow indicates a cross point. (b) A larger scale $0\leqslant\Delta\leqslant 2\varepsilon_\Lambda$.}
\label{fig:twobandSW}
\end{figure}

$D_s^\text{geo}$ and $D_s^\text{conv}$ as functions of $\Delta$ for the two-band model Eq. 1 of MT are plotted in Fig. \ref{fig:twobandSW} (same as Fig. 1 in MT), which are obtained from either numerical calculation or analytical results from Table 1 in MT. Here, we fix $\mu=0$, which means for large $\Delta/\varepsilon_\Lambda$ values, especially in Fig. \ref{fig:twobandSW}(b), the electron density will deviate from $\nu=1/2$. Fig. \ref{fig:twobandSW} resembles the $D_s$ vs $\Delta$ relation in Lieb lattice \cite{huhtinen2022revisiting}.

Notice that in Fig. \ref{fig:twobandSW}(a), the $D_s^\text{geo}$ curve for $E_g=0$ (i.e. $D_s^\text{geo}\sim \Delta\ln(\varepsilon_\Lambda/2\Delta)$) crosses the $E_g=0.1\varepsilon_\Lambda$ curve; it crosses all the $E_g\neq0$ curves, which can be seen in a larger scale in Fig. \ref{fig:twobandSW}(b) or Fig. 1 of MT. The crossing point $\Delta_c$ is of the order of $E_g$, indicating that $D_s^\text{geo}(E_g=0)>D_s^\text{geo}(E_g\neq0)$ only when $\Delta< E_g$.

This short-range property can also be understood if we fix $\Delta$ and consider the change of $D_s^\text{geo}$ vs $E_g$, as shown in Fig. 4 of MT. Then $D_s^\text{geo}$ saturates immediately as $E_g$ approaches $\Delta$. When the band gap completely closes, this saturation is necessary for a finite $D_s^\text{geo}$. Otherwise, consider the intermediate parameter regime $\Delta\ll E_g\ll W$, the $-2\ln\lambda$ piece in $f_1^\text{geo}(\lambda)$ in Table 1 of MT implies that $D_s^\text{geo}$ follows the law
\begin{align}
D_s^\text{geo}\propto\Delta\ln\frac{W}{E_g},
\end{align}
which would logarithmically diverge if $E_g\rightarrow0$. However, this logarithmic enhancement ceases and saturates as $E_g$ approaches $\Delta$.

In Fig. \ref{fig:twobandSW}(b), the $D_s$ curves for both geometric and conventional contribution reach maximum values as $\Delta$ approaches some magnitude between $E_g$ and $E_g+W$. This phenomenon is due to the inter-band reduction effect \cite{PhysRevB.109.214518}, because the usual $D_s\sim\Delta$ linear relation for flat bands cannot tend to infinity in a multi-band scenario; as $\Delta$ approaches the energy of other high-energy bands, the geometry of bands within the interaction scale gets trivialized and $D_s$ starts to decrease after the $2\Delta$ interaction window finishes including the high-energy bands. This geometry trivialization also explains in Fig. 4 of MT why $D_s^\text{geo}$ decreases slightly when $E_g$ keeps closing after reaching $\Delta$.
\section{S5: Uniform Pairing Condition at Zero-temperature for Chiral Models}
\label{app:upc}
This section explains the uniform pairing condition (UPC) used to derive the superfluid weight formula in this paper. We want to emphasize that for a general tight-binding model, the UPC can always be strictly satisfied for intra-orbital interactions by tuning the attractions $U_\alpha$. However, if $U_\alpha$s are fixed, then the UPC is only an approximation in some cases.

Take the bipartite Lieb lattice model Eq. \eqref{eq:lieb} as an example. There are two limits: (1) the gap-closed limit $t\delta\ll \Delta$ and (2) the isolated limit $t\delta\gg\Delta$. For the former, although the flat band has no component from orbital $B$ (see Fig. \ref{fig:lieblattice}(b)), the two dispersive bands that have entered the interaction scale do contain orbital $B$. Therefore, if one wants to impose UPC and correctly consider the inter-band effect between the two dispersive bands, $U_B/U_A$ must be tuned as $\delta$ changes. When $\delta$ increases from 0 to $\delta\sim \Delta/t$, the $B$-orbital-resolved density of states in the interaction window vanishes, implying $U_B\rightarrow\infty$. This shows that in the regime $0\leq\delta\lesssim \Delta/t$, UPC is not a good approximation if one fixes $U_B$. However, once the isolated limit is reached, orbital $B$ has been removed from the interaction scale, then the value of $U_B$ will not affect the superfluid weight. Therefore, in the regime $t\delta\gg\Delta$, UPC is always well established as long as $U_A=U_C$, as pointed out by Ref. \cite{julku2016geometric}.

Here we solve the self-consistency gap equations for the chiral models Eq. 1 and 3 of MT, to point out in what parameter regime the UPC is a good approximation if $U_\alpha$ are fixed. Unlike the bipartite Lieb lattice, in the three-band chiral model, as the band gap is opened by the mass $m$, some orbital $B$ component shifts from the dispersive bands to the flat band. Therefore, it is never removed from the interaction scale.

\begin{figure}[t!]
\includegraphics[width=0.96\textwidth]{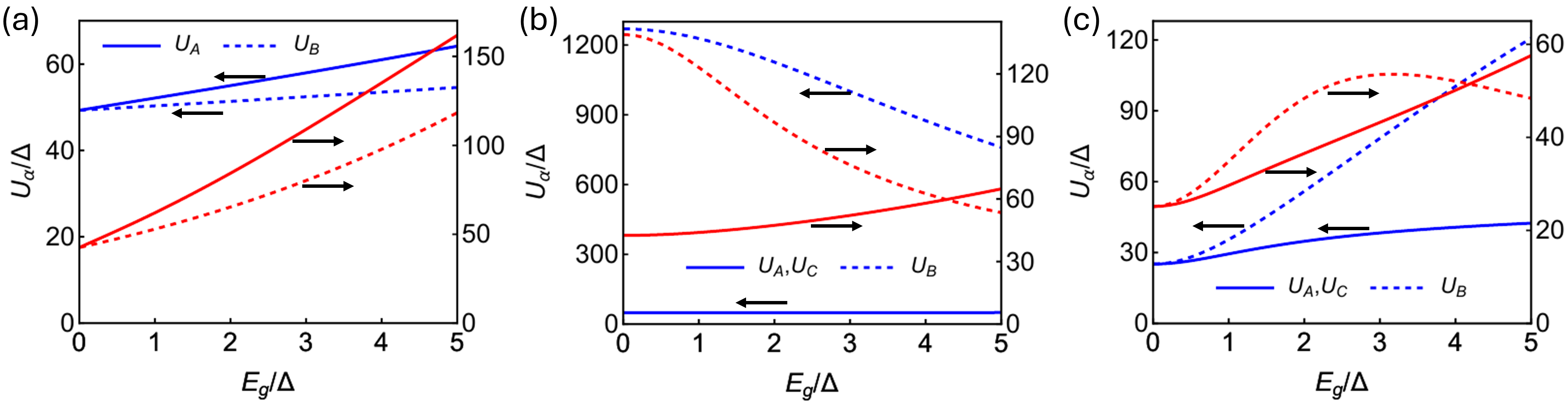}
\caption{$U_\alpha$ as function of $E_g$ solved from UPC $\Delta_\alpha\equiv\Delta$ for the chiral models with $J=\zeta_J=k_\Lambda=1$. Blue curves: $\Delta=0.01\varepsilon_\Lambda$; Red curves: $\Delta=0.1\varepsilon_\Lambda$. (a) for two-band model Eq. 1 of MT, $E_g=2m$, $\varepsilon_\Lambda=2\zeta_Jk_\Lambda^J$; (b)(c) for case (i),(ii) of the three-band model Eq. 3 of MT, respectively, with $E_g=m$, $\varepsilon_\Lambda=\zeta_Jk_\Lambda^J$.}
\label{fig:ugap}
\end{figure}

In Fig. \ref{fig:ugap}, we impose UPC by fixing all $\Delta_\alpha\equiv\Delta$ and solve $U_\alpha$ as a function of $E_g$ from the gap equation:
\begin{align}
1=\frac{U_\alpha}{N}\sum_{\bd{k},l}\frac{|u_{l\bd{k}\alpha}|^2}{2E_{l\bd{k}}},
\end{align}
Where $\bd{k}$ is summed over the cutoff region $k<k_f=1$, $l$ is summed over all the two or three bands of the model, and $\mu$ is set to be 0. For the two-band and three-band chiral model, \tit{the variation of the ratio $U_A/U_B$ with $E_g$ tells whether UPC is a good approximation in some regime} (for the three-band model we set $U_A=U_C$). In Fig. \ref{fig:ugap}(a), we have $U_A=U_B$ at $E_g=0$ because the two orbitals are symmetric; as the gap opens, $U_A, U_B$ increase because the density of both orbitals near $\mu=0$ are lowered. Compared to the $\Delta=0.1\varepsilon_\Lambda$ case (red curves), we find the change of $U_A/U_B$ with $E_g$ is smaller for $\Delta=0.01\varepsilon_\Lambda$ (blue curves). Therefore, for the same regime $0<E_g/\Delta<5$, UPC is a better approximation for smaller $\Delta$. Fig. \ref{fig:ugap}(b)(c) can be similarly analyzed. We find for case (i) of the three-band model ($C(m,k)=1$, Fig. \ref{fig:ugap}(b)), UPC is again better for the smaller $\Delta$, whereas for case (ii) of the three-band model with $c(m)=m$ (Fig. \ref{fig:ugap}(c)), UPC is better for larger $\Delta$ (red curves) in the entire regime $0<E_g/\Delta<5$.
\section{S6: Stability of the Uniform Pairing Condition with Temperature}
\label{app:stability}
We analytically show the uniform pairing condition (UPC) for flat bands in proximity to dispersive bands is stable against the change of temperature if the interaction is weak (i.e., either $\Delta<E_g$ or $\Delta>E_g$ but $\Delta-E_g\ll W_d$, both of which lead to much smaller electron density in the dispersive bands than the flat bands).

The gap equation for an $s$-orbital superconductor with onsite intra-orbital density-density interaction, $\Delta_\alpha(T)=-U_\alpha\la c_{i\alpha\downarrow}c_{i\alpha\uparrow}\ra\,(1\leqslant\alpha\leqslant s)$ can be expressed as
\begin{align}\label{eq:gap}
\Delta_\alpha(T)=-\frac{U_\alpha}{N}\sum_{\bd{k},ll'}u_{l\bd{k}\alpha}^*u_{l'\bd{k}\alpha}\la c_{l,-\bd{k}\downarrow}c_{l'\bd{k}\uparrow}\ra,
\end{align}
where $c_{l\bd{k}\sigma}$ is the electron operator in band basis and the band indices $l,l'$ sum over all relevant bands. $u_{l\bd{k}\alpha}$ are the Bloch function components and $\la\ra$ is the thermal average over the superconducting state. Eq. \eqref{eq:gap} is a set of $s$ equations that solve for the $s$ variables $\Delta_\alpha$, and these $\Delta_\alpha$ are coupled nonlinearly. \tit{We assume the flat bands contain all these $s$ orbitals}. For fixed values of temperature $T$ and interactions $U_\alpha$, a solution $\Delta_1,...,\Delta_s$ in general exists. However, the uniform solution $\Delta_1=...=\Delta_s$ may exist for some special parameters only if there is no symmetry between orbitals. e.g. it can always be achieved at $T=0$ for fixed model parameters (so $W$ and $E_g$ are fixed), by adjusting $U_\alpha$ values, but then it will not strictly hold for $T\neq0$. We want to study whether the uniform solution can persist at finite temperatures.

Let's choose parameters $U_\alpha$ such that UPC is satisfied at $T=0$; we further make the \tit{assumption} that it also holds for all finite $T$ below the transition temperature, i.e., $\Delta_1(T)=...=\Delta_s(T)=\Delta(T)$ (this assumption will be checked in the following calculations). With this assumption, Eq. \eqref{eq:gap} simplifies to
\begin{align}\label{eq:gap2}
1=\frac{U_\alpha}{N}\sum_{\bd{k},l}\frac{|u_{l\bd{k}\alpha}|^2}{2E_{l\bd{k}}}\tanh\frac{\beta E_{l\bd{k}}}{2},
\end{align}
where $E_{l\bd{k}}=\sqrt{\xi_{l\bd{k}}^2+\Delta(T)^2}$. Now that the $s$ equations of Eq. \eqref{eq:gap2} are decoupled, solving $\Delta(T)$ from them individually gives a curve $\widetilde{\Delta}_\alpha(T)$. Then, our assumption can be proved if these $s$ curves coincide.

\begin{figure}[b!]
\includegraphics[width=0.48\textwidth]{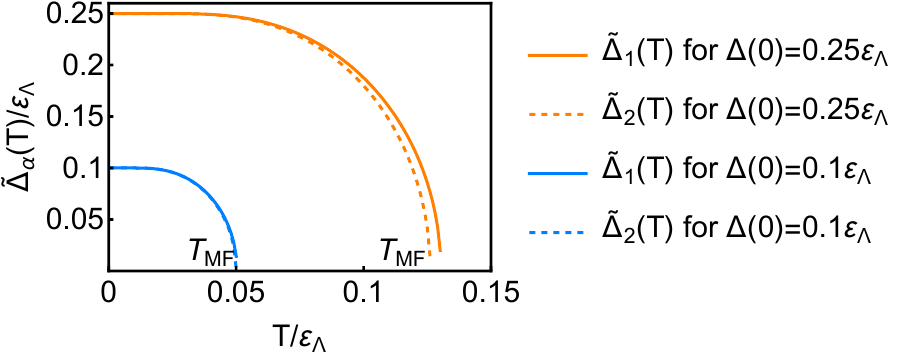}
\caption{Plot of $\widetilde{\Delta}_\alpha(T)$ ($\alpha=1,2$) solved from Eq. \eqref{eq:gap2} for the two-band chiral model Eq. 1 of MT with $J=1$, $\zeta_J=1$. $m$ is fixed to be $0.2W_d$ so $E_g=2m=0.4W_d$. Blue curves for the case $\Delta(0)<E_g$; orange curves for the case $\Delta(0)>E_g$ but $\Delta(0)-E_g\ll W_d$.}
\label{fig:twogap}
\end{figure}

Choosing any $\alpha$th equation from Eq. \eqref{eq:gap2}, it reads
\begin{align}\label{eq:alphaeq}
1=\frac{U_\alpha}{N}\sum_\bd{k}\bigg(\sum_{l\in\text{flat}}\frac{|u_{l\bd{k}\alpha}|^2}{2E_{l\bd{k}}}\tanh \frac{\beta E_{l\bd{k}}}{2} + \sum_{l'\in\text{disp}}\frac{|u_{l'\bd{k}\alpha}|^2}{2E_{l'\bd{k}}}\tanh \frac{\beta E_{l'\bd{k}}}{2}\bigg),
\end{align}
where $l$ is summed over all the degenerate flat bands and $l'$ is over all dispersive bands. Under the condition $\mu\approx 0$ and $\Delta(0)<E_g$ or $\Delta(0)>E_g$ but $\Delta(0)-E_g\ll W_d$, the r.h.s. of Eq. \eqref{eq:alphaeq} is mainly contributed by the first term because these from the dispersive bands are cut off by factor $1/E_{l'\bd{k}}$ in most regions of $\bd{k}$ space. This means Eq. \eqref{eq:alphaeq} can be approximated as
\begin{align}\label{eq:alphaeq2}
1\approx\frac{U_\alpha}{N}\sum_{\bd{k},l\in\text{flat}}\frac{|u_{l\bd{k}\alpha}|^2}{2E_{l\bd{k}}}\tanh \frac{\beta E_{l\bd{k}}}{2},\,\,\,\forall T.
\end{align}
Combining the Eq. \eqref{eq:alphaeq2} at $T=0$ and $T=T_{MF}$ (defined from $\widetilde{\Delta}_\alpha(T_{MF})=0$), it leads to
\begin{align}
\sum_{\bd{k},l\in\text{flat}}|u_{l\bd{k}\alpha}|^2\bigg[\frac{1}{2T_{MF}}-\frac{1}{\Delta(0)}\bigg]\approx 0,
\end{align}
which gives $T_{MF}\approx 0.5\Delta(0)$ regardless of orbital index $\alpha$. Since we enforced UPC at $T=0$, $\widetilde{\Delta}_1(0)=...=\widetilde{\Delta}_s(0)=\Delta(0)$, this result shows that they have the same $T_{MF}$. We conclude that for all the orbitals $\alpha$ that contribute to the flat bands, their $\widetilde{\Delta}_\alpha(T)$ curve almost coincide so that the UPC can persist to any finite temperature. On the other hand, if an orbital mainly contributes to the dispersive bands, then its UPC will not persist to finite temperature even if it is satisfied at $T=0$.

In Fig. \ref{fig:twogap}, we calculated the $\widetilde{\Delta}_\alpha(T)$ curves for the two-band chiral model. For this model, if the band gap completely closes ($m=0$), then the two orbitals are symmetric, and $\Delta_1(T)=\Delta_2(T)$ holds for any temperature once we set $U_1=U_2$. Therefore, we are interested in the case $m\neq0$. As long as the mass $m$ is not too large, both orbitals $\alpha=1,2$ contribute to the flat band, so the curves show the UPC holds well until $T$ gets very close to $T_{MF}$.

\section{S7: Temperature Dependence of Superfluid Weight and BKT Transition}
\label{app:temperature}
For finite temperatures we solve a simplified gap equation for $\Delta_\alpha$ with parameters $E_g$ and $T$:
\be\label{eq:gapapprox}
\Delta_\alpha=\frac{U_\alpha}{N}\sum_{\bd{k},l}|u_{l\bd{k}\alpha}|^2\frac{\bar{\Delta}}{2E_{l\bd{k}}}\tanh\frac{\beta E_{l\bd{k}}}{2}+O(\delta\Delta_\alpha),
\ee
where $\bar{\Delta}=(1/s)\sum_{\alpha=1}^s\Delta_\alpha$ ($s=2$ or 3) is the averaged order parameter, $\delta\Delta_\alpha=\Delta_\alpha-\bar{\Delta}$ and $E_{l\bd{k}}=\sqrt{\xi_{l\bd{k}}^2+\bar{\Delta}^2}$. Eq. \eqref{eq:gapapprox} is a good approximation of the exact equation \eqref{eq:gap} when the pairing matrix $\hat{\Delta}=\text{diag}\{\Delta_1,...,\Delta_s\}$ is close to uniform, with the inter-band $O(\delta\Delta_\alpha)$ terms neglected. Its solution tells the lowest order how $\hat{\Delta}$ deviates from uniform pairing.

\begin{figure}[h!]
\begin{center}
\includegraphics[width=0.96\textwidth]{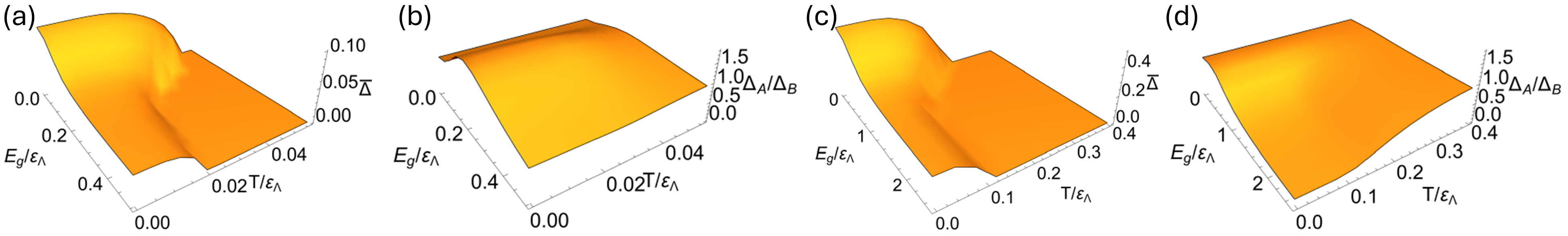}
\caption{$\bar{\Delta}$ and $\Delta_A/\Delta_B$ solved for the flattened three-band model, with $c(m)=m$. (a)(b) Fixing $U_A=U_B=2.51\varepsilon_\Lambda$ such that UPC is satisfied at $E_g=T=0$, with $\Delta_\alpha(0,0)=0.1\varepsilon_\Lambda$; (c)(d) similarly, fixing $U_A=U_B=12.56\varepsilon_\Lambda$ such that $\Delta_\alpha(0,0)=0.5\varepsilon_\Lambda$.}
\label{fig:3dplot}
\end{center}
\end{figure}

\begin{figure}[h!]
\begin{center}
\includegraphics[width=0.48\textwidth]{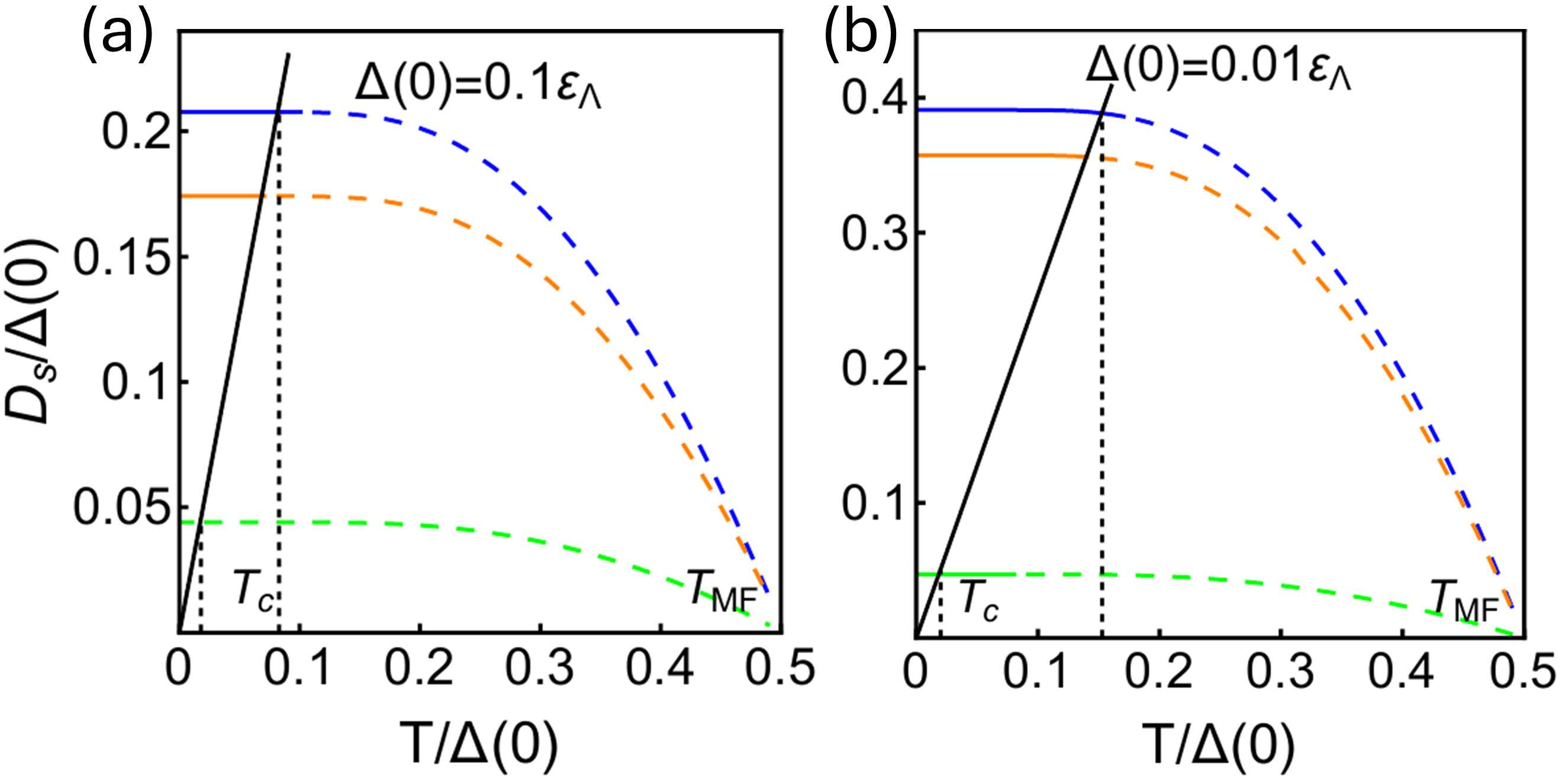}
\caption{$D_s$ vs temperature for the two-band model Eq. 1 of MT. Different colors show $E_g=0.5\varepsilon_\Lambda$ (green), $E_g=\Delta(0)$ (orange) and $E_g=0$ (blue). (a) $\Delta(0)=0.1\varepsilon_\Lambda$, (b) $\Delta(0)=0.01\varepsilon_\Lambda$.}
\label{fig:dsvstplot}
\end{center}
\end{figure}

As an example, in Fig. \ref{fig:3dplot}, we plot the $\Delta_\alpha(E_g,T)$ solved from Eq. \eqref{eq:gapapprox} for the flattened three-band chiral model (case (ii)) by taking $c(m)=m$. In the calculation, $U_A,U_B$ are fixed such that the UPC is satisfied at $E_g=0, T=0$. In Fig. \ref{fig:3dplot}(a)(b), the UPC $\Delta_\alpha(0,0)=0.1\varepsilon_\Lambda$, we find in (a) as $E_g=m$ closes, $\bar{\Delta}$ increases due to the enhanced density of states, so $T_{MF}$ (the temperature at which $\bar{\Delta}$ decrease to 0) also increases. (b) shows the ratio of $\Delta_A/\Delta_B$ which tells how uniform the pairing is. At $T=0$, as $m$ increases, the pairing becomes nonuniform at $E_g/\varepsilon_\Lambda=0.2$ (i.e., $E_g/\bar{\Delta}\approx 2$) and becomes uniform again at $E_g/\varepsilon_\Lambda=0.4$ ($E_g/\bar{\Delta}\approx 4$), which reproduces the red curves in Fig. \ref{fig:ugap}(c). This non-uniformity with $E_g$ becomes less severe as $T$ increases. In Fig. \ref{fig:3dplot}(c)(d) we show the solutions with UPC $\Delta_\alpha(0,0)=0.5\varepsilon_\Lambda$ (which was not shown in Fig. \ref{fig:ugap}(c)). In (d), we find the UPC holds well except for the regime $E_g/\varepsilon_\Lambda>1.5$, $T/\varepsilon_\Lambda<0.2$, as the model approaches the atomic limit.

After obtaining $\Delta_\alpha(T)$, we use UPC as an approximation and insert $\bar{\Delta}(T)$ into Eq. \eqref{eq:dst} to compute $D_s(T)$. From this, one can further calculate the BKT transition temperature $T_c$, as shown in Fig. 4(b) and Fig. 5(b) of MT. Fig. \ref{fig:dsvstplot} shows the temperature-dependence of $D_s$ using the example of the two-band chiral model. The $D_s(T)$ curves are very flat at $T<0.3T_{MF}$ and have a finite slope near $T_{MF}$, which is a general feature of flat bands with $s$-wave uniform order parameter. $T_c$ is extracted from the $D_s(T)$ curves by their intersections with the universal line of $(8/\pi)T$. When $D_s(0)/\Delta(0)$ is small, the intersections lie in the flat part of the curves and give $T_c\approx (\pi/8)D_s(0)$, exhibiting a Uemura relation \cite{uemura1989universal}.

\section{S8: $D_s/\Delta$ vs $E_g/\Delta$ scaling plots and $T_c$ vs $E_g$ plots for large $J$ Values}
\label{app:higherj}
\begin{figure}[h!]
\begin{center}
\includegraphics[width=0.75\textwidth]{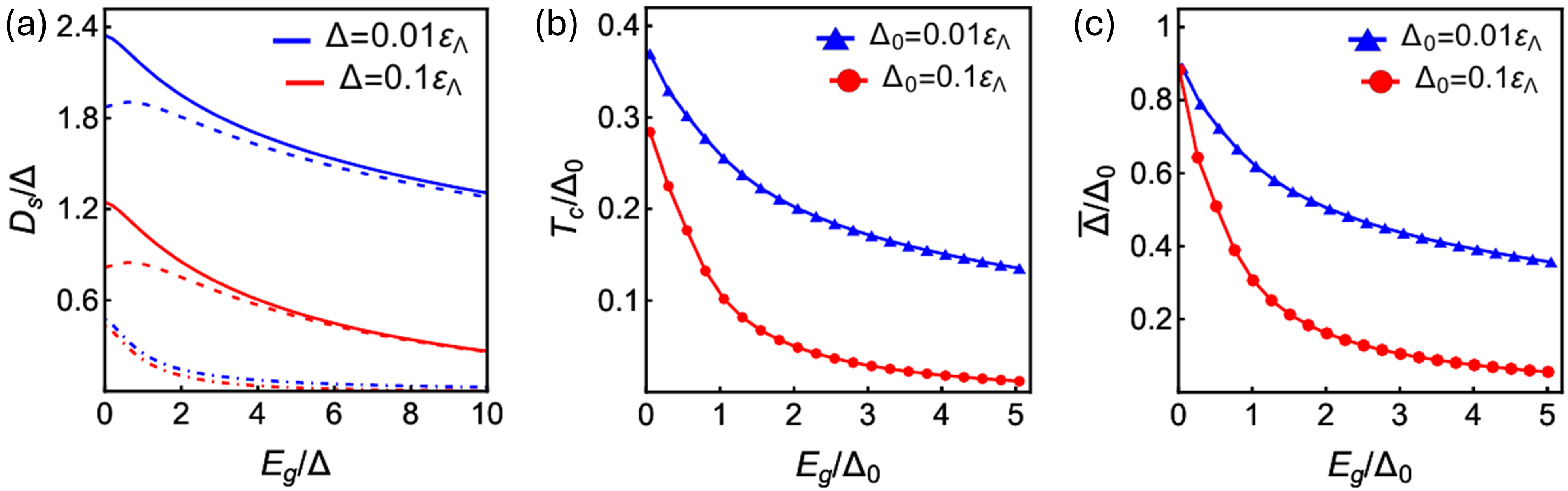}
\caption{The $D_s/\Delta$ vs $E_g/\Delta$ scaling behavior, $T_c$ and the averaged uniform pairing $\bar{\Delta}$ vs $E_g$ for the two-band model (Eq. 1 of MT), with $J=6$. The conventions are the same as for Fig. 4 of the MT. In (b), $U_{A,B}$ are fixed at $0.78\varepsilon_\Lambda$ (blue) and $6.28\varepsilon_\Lambda$ (red) for the two weak couplings, respectively. (c) is the zero-temperature averaged uniform pairing $\bar{\Delta}$ vs $E_g$. The magnitude of $\bar{\Delta}$ increases a lot due to the large density of states at the edge of the dispersive band of $J=6$.}
\label{fig:2bandj6}
\end{center}
\end{figure}

\begin{figure}[h!]
\begin{center}
\includegraphics[width=0.75\textwidth]{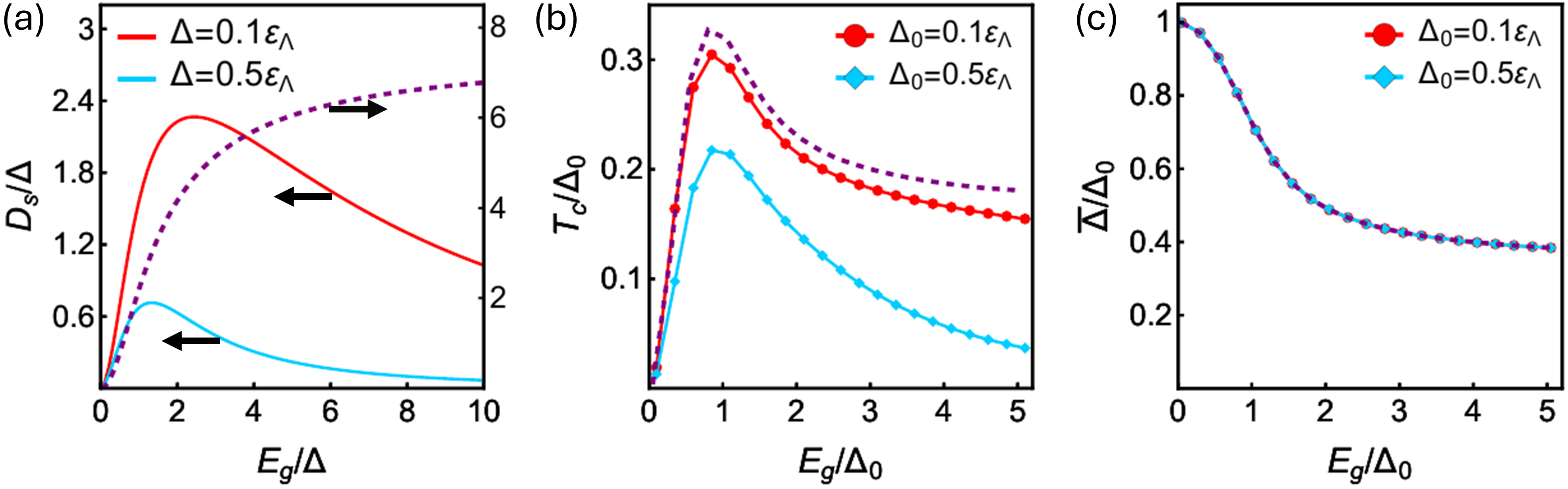}
\caption{The $D_s/\Delta$ vs $E_g/\Delta$ scaling behavior, $T_c$ and the averaged uniform pairing $\bar{\Delta}$ vs $E_g$ for the flattened three-band model (Eq. 3 of MT, case (ii)), with $J=6$. The conventions are the same as for Fig. 5 of the MT, except here we fix $m=m_0=0.032\varepsilon_\Lambda$. In (b), $U_{A,B,C}=2.51\varepsilon_\Lambda$ (red) and $12.56\varepsilon_\Lambda$ (light blue), while the non-singular model curve (dashed purple) is independent of the interactions. In (c), $\bar{\Delta}$ of the three cases ($\Delta_0=0.1\varepsilon_\Lambda$, $\Delta_0=0.5\varepsilon_\Lambda$ and non-singular) coincide.}
\label{fig:3bandfj6}
\end{center}
\end{figure}

As an example, we discuss these results for the representative value $J=6$. The two cases of weak and strong interaction limits are shown in Fig. \ref{fig:2bandj6} and \ref{fig:3bandfj6}, respectively. In both Fig. \ref{fig:2bandj6}(a) and \ref{fig:3bandfj6}(a), $D_s/\Delta$ scales with $J$, except for the non-singular model (the dashed curve in Fig. \ref{fig:3bandfj6}(a)). For the flattened three-band non-singular case, the choice $m_0$ that makes the uniform pairing best satisfied depends on $J$; for $J=6$, $m_0=0.032\varepsilon_\Lambda$, which makes $D_s/\Delta$ of this curve not scale with $J$ directly.

The large $J$ value makes $D_s/\Delta>1$ in most regimes of Fig. \ref{fig:2bandj6}(a) and \ref{fig:3bandfj6}(a), therefore, the $T_c$ enhancement due to the gap closing is mainly contributed by the superconducting gap $\Delta$. For the weak interaction case, one can see the $T_c$ enhancement by gap closing for $J=6$ (Fig. \ref{fig:2bandj6}(b)) is stronger than for $J=1$ (Fig. 4(b) in the MT). The reason is that the edge of the dispersive band of $J=6$ is much flatter than $J=1$, leading to a large increase of the density of states, which further enhances $\Delta$. In order to see for large $J$ the $T_c$ is mainly enhanced by $\Delta$ rather than $D_s$, the best way is to plot the zero-temperature $\Delta$ vs $E_g$, as shown in Fig. \ref{fig:2bandj6}(c), and compare it with the $T_c$ curves in Fig. \ref{fig:2bandj6}(b). We indeed find they resemble each other. An alternative way is by noticing that $T_c/\Delta_0$ in Fig. \ref{fig:2bandj6}(b) is close to $0.5$ as the gap completely closes, since the saturation value is $T_{MF}/\Delta_0=0.5$ for flat bands (in the limit $D_s\gg\Delta$, the BKT transition temperature $T_c=T_{MF}$).

Similarly, in the strong interaction case, we see the $T_c$ enhancement of the singular model is no stronger than the non-singular model (Fig. \ref{fig:3bandfj6}(b)), since the enhancement of $\Delta$ by gap closing in both cases are the same (Fig. \ref{fig:3bandfj6}(c)). The three curves in Fig. \ref{fig:2bandj6}(b) are slightly different because their distinct $D_s/\Delta$ behaviors (Fig. \ref{fig:3bandfj6}(a)). The $E_g/\Delta_0>1$ regime of the curves in Fig. \ref{fig:2bandj6}(b) also resemble the $\Delta$ vs $E_g$ curves in Fig. \ref{fig:3bandfj6}(c), confirming that the $T_c$ enhancement is by $\Delta$ rather than $D_s$.

\end{document}